\documentclass[aip,apl,reprint,amsmath,amssymb]{revtex4-1}

\usepackage{amsmath}
\usepackage{wasysym}
\usepackage{upgreek} 
\usepackage{graphicx}

\begin{document} 

\title{Laser radio transmitter} 

\author{Marco Piccardo}
\email[]{These authors contributed equally to this work.}
\affiliation{Harvard John A. Paulson School of Engineering and Applied Sciences, Harvard University, Cambridge, MA 02138 USA}

\author{Michele Tamagnone}
\email[]{These authors contributed equally to this work.}
\affiliation{Harvard John A. Paulson School of Engineering and Applied Sciences, Harvard University, Cambridge, MA 02138 USA}

\author{Benedikt Schwarz}
\affiliation{Harvard John A. Paulson School of Engineering and Applied Sciences, Harvard University, Cambridge, MA 02138 USA}
\affiliation{Institute of Solid State Electronics, TU Wien, 1040 Vienna, Austria}

\author{Paul Chevalier}
\affiliation{Harvard John A. Paulson School of Engineering and Applied Sciences, Harvard University, Cambridge, MA 02138 USA}

\author{Noah A. Rubin}
\affiliation{Harvard John A. Paulson School of Engineering and Applied Sciences, Harvard University, Cambridge, MA 02138 USA}

\author{Yongrui Wang}
\affiliation{Department of Physics and Astronomy, Texas A\&M University, College Station, TX 77843, USA}

\author{Christine A. Wang}
\affiliation{Lincoln Laboratory, Massachusetts Institute of Technology, Lexington, Massachusetts 02420, USA}

\author{Michael K. Connors}
\affiliation{Lincoln Laboratory, Massachusetts Institute of Technology, Lexington, Massachusetts 02420, USA}

\author{Daniel McNulty}
\affiliation{Lincoln Laboratory, Massachusetts Institute of Technology, Lexington, Massachusetts 02420, USA}

\author{Alexey Belyanin}
\affiliation{Department of Physics and Astronomy, Texas A\&M University, College Station, TX 77843, USA}

\author{Federico Capasso}
\email[]{capasso@seas.harvard.edu}
\affiliation{Harvard John A. Paulson School of Engineering and Applied Sciences, Harvard University, Cambridge, MA 02138 USA}

\date{\today}

\begin{abstract}
Since the days of Hertz, radio transmitters have evolved from rudimentary circuits emitting around 50 MHz to modern ubiquitous Wi-Fi devices operating at gigahertz radio bands. As wireless data traffic continues to increase there is a need for new communication technologies capable of high-frequency operation for high-speed data transfer. Here we give a proof of concept of a new compact radio frequency transmitter based on a semiconductor laser frequency comb. In this laser, the beating among the coherent modes oscillating inside the cavity generates a radio frequency current, which couples to the electrodes of the device. We show that redesigning the top contact of the laser allows one to exploit the internal oscillatory current to drive an integrated dipole antenna, which radiates into free space. In addition, direct modulation of the laser current permits encoding a signal in the radiated radio frequency carrier. Working in the opposite direction, the antenna can receive an external radio frequency signal, couple it to the active region and injection lock the laser. These results pave the way to new applications and functionality in optical frequency combs, such as wireless radio communication and wireless synchronization to a reference source.
\end{abstract}

\maketitle 

\begin{figure*}[t]
\centering
\includegraphics[width=0.95\textwidth]{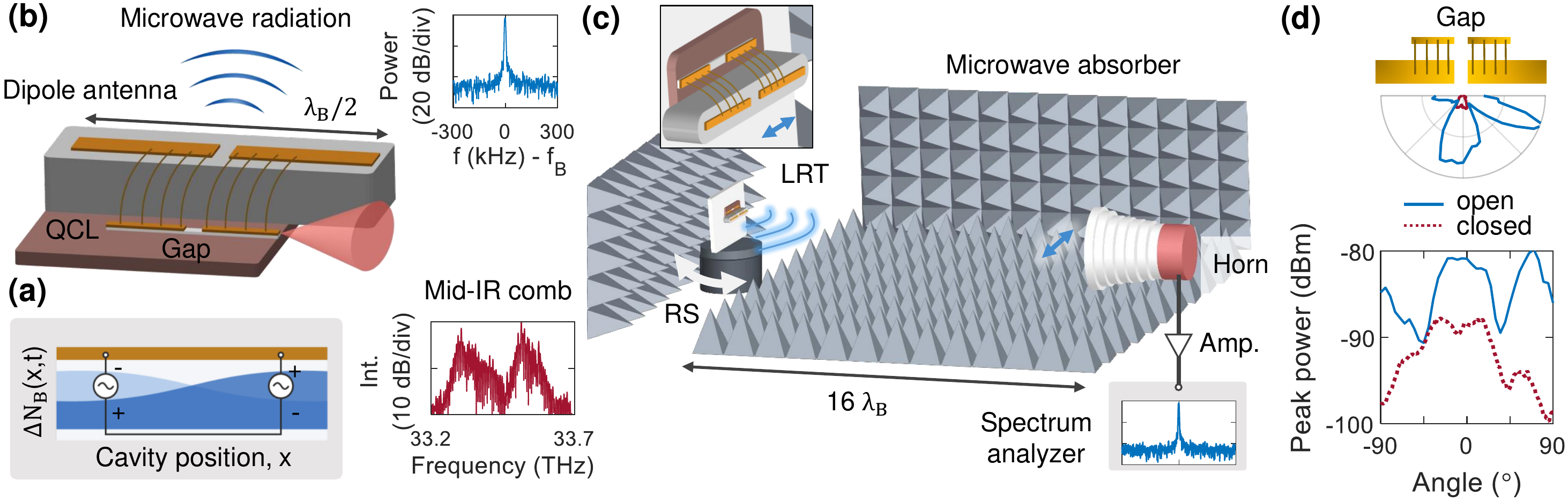}
\caption{\textbf{(a)} Schematic of the time-dependent population inversion grating oscillating inside the cavity of a quantum cascade laser frequency comb at the fundamental beat note frequency. In light of this phenomenon the laser can be regarded as two radio frequency generators oscillating in antiphase. \textbf{(b)} Introducing a gap in the design of the top electrode of the device allows one to use the radio frequency alternating currents generated inside the laser to feed a dipole antenna enabling wireless microwave emission, in addition to the usual mid-infrared radiation. The insets show the microwave beat note emitted from the device (top, $f_B=5.5$~GHz) and the mid-infrared frequency comb spectrum (bottom). \textbf{(c)} Set-up for the characterization of the far-field pattern. The laser radio transmitter (LRT) is mounted on a rotation stage (RS) and the 5.5~GHz radiation emitted at different angles in the horizontal plane is measured at a distance of 0.9~m by a horn antenna connected to a spectrum analyzer. $\lambda_B$ denotes the free space wavelength of the radiated beat note. Microwave absorbers are used to eliminate unwanted reflections from the surrounding environment. The inset shows a magnification of the LRT with the polarization of the emitted microwave field (double-headed arrow), which is the same one of the receiving horn antenna. \textbf{(d)} The radiation patterns are measured in the case where the gaps of the QCL and dipole antenna are open and closed by wirebonds. The direction normal to the surface of the LRT defines the 0$^{\circ}$ angle. Polar plots are shown with power plotted on a linear scale. The amplifier gain has been subtracted.}
\label{fig1_conceptwireless}
\end{figure*}

Optical fields can be used to synthesize low-phase-noise microwaves by means of different techniques, such as optical frequency division~\cite{Fortier2011,Li309}, optoelectronic oscillations~\cite{Maleki2011} and laser heterodyne~\cite{Li2013}. The latter can be realized in a medium with a nonlinear optical response, such as a photomixer~\cite{Tani2005}, capable of converting the frequency difference between the optical modes into a microwave tone~\cite{Kim:10,Rubiola2013,Quinlan2013}. An attractive aspect of quantum cascade lasers (QCLs) operating as optical frequency combs~\cite{Udem2002, Hansch2006, Hugi2012,Villares2014,Burghoff2014,Lu2015,Cappelli2016} is that they can act both as sources of light whose spectrum consists of equidistant modes and as photomixers --- provided that their gain dynamics are sufficiently fast --- producing microwaves of high spectral purity directly inside the laser cavity. The physical process underlying the microwave generation consists in the beating among neighboring optical modes of the standing wave cavity, which induces a dynamic electronic grating with opposite ends oscillating nearly in antiphase at the beat frequency (Fig.~\ref{fig1_conceptwireless}a). Such mechanism was used to demonstrate a quadrature modulation scheme exploiting the alternating currents oscillating inside the laser and a system of near-field microwave probes~\cite{Piccardo:18}.

In light of this phenomenon the laser can be viewed from a new perspective, namely as an ensemble of two radio frequency generators with opposite phase. Usually the top electrode of these lasers consists of an electrically continuous metal contact connecting the two generators and thus preventing the device from radiating. In this work we demonstrate that by adapting the geometry of the top contact layer of a QCL allows us to feed a dipole antenna integrated on the chip, enabling emission of radio waves into free space. The beatnote frequency itself can be tuned by modulating the laser current, thus the laser acquires a new functionality, turning into a radio transmitter capable of wireless communication at a carrier frequency of 5.5 GHz, given by the comb repetition rate. Thanks to their fast gain recovery dynamics~\cite{Choi2008PRL}, QCLs have the potential of generating sub-terahertz carriers when operating in the harmonic comb regime with a wide intermodal spacing~\cite{Kazakov2017}. The extension of the design presented here to realize a new class of terahertz wireless communication devices~\cite{Nagatsuma2016,Akyildiz2014,Kleine-Ostmann2011} will be discussed.

The laser radio transmitter (LRT) is illustrated in Fig.~\ref{fig1_conceptwireless}b. It consists of a continuous wave, ridge waveguide, uncoated Fabry-Perot QCL with an 8~mm-long cavity operating as a fundamental frequency comb~\cite{Hugi2012} in the mid-infrared spectral range with a narrow (kilohertz) linewidth beat note at $f_B=5.5$~GHz (Fig.~\ref{fig1_conceptwireless}b). A 400~$\mu$m wide gap is etched in the top contact layers of the device creating two contact sections with an open-circuit resistance of 250~$\Omega$. The two top laser contacts are connected through wirebonds to a low-impedance half-wave dipole antenna designed to radiate at $f_B$ consisting of two gold stripes fixed on a polyactide dielectric substrate. The laser current is injected from the DC power source through the antenna into the QCL (Methods).

To characterize the emission pattern of the system we carry out far-field measurements of the QCL microwave radiation. A schematic of the experimental set-up is shown in Fig.~\ref{fig1_conceptwireless}c. The LRT is mounted on a rotation stage and the microwave radiation emitted at angles between $-90^\circ$ and $90^\circ$ on the horizontal plane is detected by a directive horn antenna (gain 18.5 dBi) at a distance of 0.9~m (16~$\lambda_B$) from the source, amplified and then measured with a spectrum analyzer (Supplementary Sec.~1). Fig.~\ref{fig1_conceptwireless}d shows the measured radiation pattern of the device (continuous line). The central peak emitted around the normal to the surface of the LRT originates from the dipole antenna and QCL, while the side lobes are due to emission from the wirebonds (Supplementary Sec.~2). The maximum radiated power is approximately $-80$ dBm. When the gap of the QCL and antenna is closed using wirebonds (Fig.~\ref{fig1_conceptwireless}d, dashed line), the maximum power drops to $-88$~dBm. Even though this does not correspond to the case of an unstructured device with a continuous metal electrode --- because of the inductive nature of the wirebonds which prevents them from being a perfect short --- this result shows the fundamental role that the gap geometry plays for the wireless emission.

From the measured microwave radiation, taking into account the directivity of the emitter and receiver antenna, and using an equivalent circuit model of the LRT, we can estimate the microwave power available at the source. The QCL active region can be modeled as a radio frequency generator with low output impedance. The bonding pads on the sides of the waveguide behave as capacitors in parallel to the generator. A full model of the impedance of the pads and of the other elements in the equivalent circuit is described in Supplementary Sec.~3. The antenna is connected to the QCL via wirebonds, whose inductive behavior at microwave frequencies is also taken into account in the model. Finally, the antenna constitutes the load connected at the other end of the circuit. The impedance mismatch loss (mostly caused by the presence of the capacitive pads) is estimated to be $-22$~dB, which means that the power actually radiated by the antenna is 22 dB lower than the available power at the QCL. Using this fact and knowing that the total received power is $-80$~dBm, it is possible to compute the wireless link power budget with the Friis formula (Supplementary Sec.~3). The total radiated power is found to be $-58$~dBm, meaning that the available power at the QCL is estimated to be $-36$~dBm. Based on numerical simulations of the QCL radio frequency generation, the available power is expected to increase by several orders of magnitude when operating the laser in the harmonic comb regime, due to the higher optical power per mode and lower number of beat harmonics that are generated in this state (Supplementary Sec.~4). On the other hand, the source impedance mismatch due to the pad admittance can be corrected using a buried heterostructure geometry with a Fe-doped InP insulating layer~\cite{Nida2017}, promising to improve the extraction performances both at microwave and sub-terahertz frequencies, as the capacitive pad admittance would be substantially reduced. \\

\begin{figure}[t]
\centering
\includegraphics[width=0.48\textwidth]{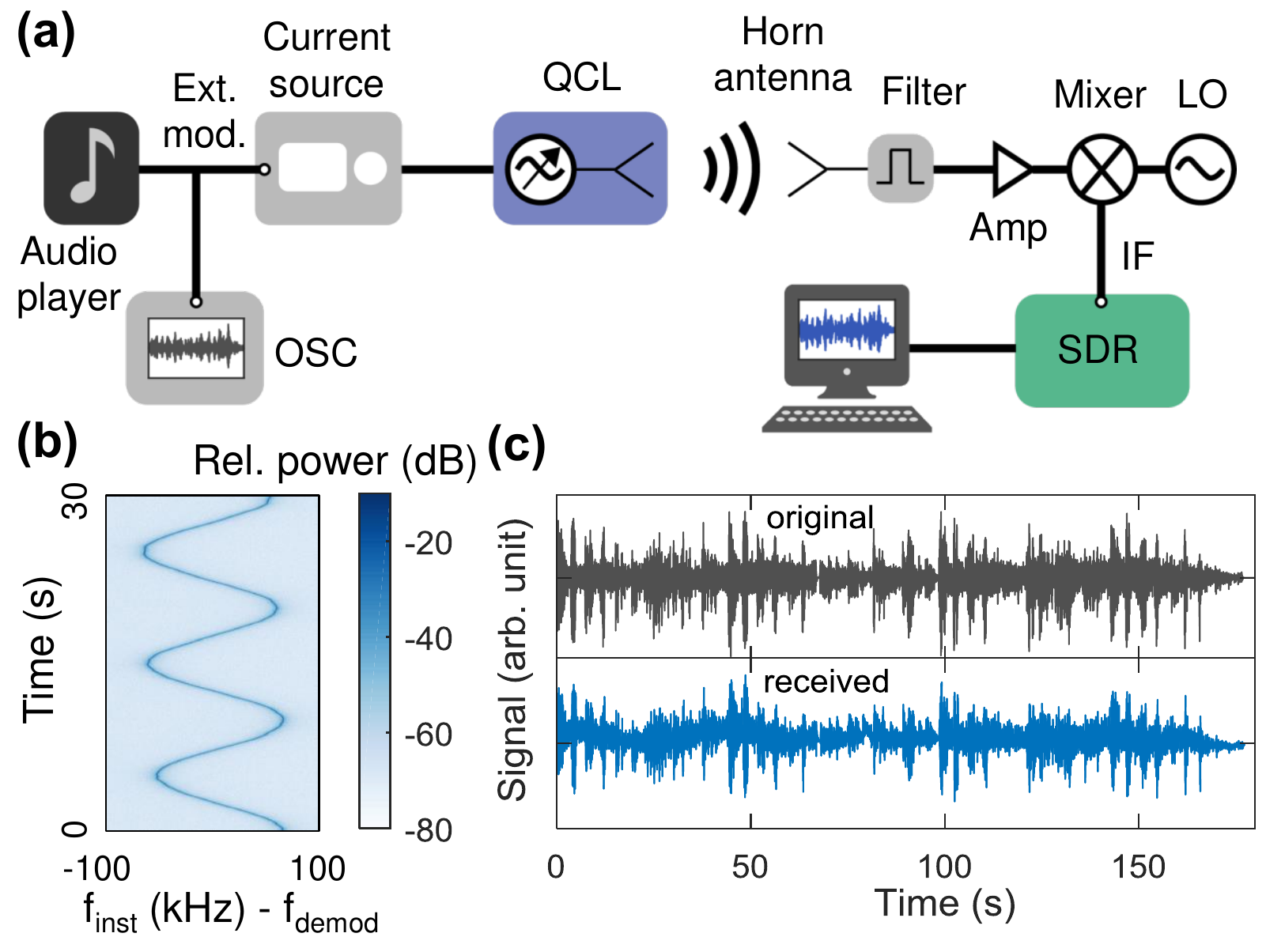}
\caption{\textbf{(a)} Set-up for the wireless transmission and reception of an audio signal. The laser current is modulated by an analog signal inducing a frequency modulation of the laser beat note. The radio signal is received by a horn antenna, filtered and downconverted to fit into the bandwidth of a software-defined radio (SDR). LO, local oscillator, IF intermediate frequency, OSC, oscilloscope. \textbf{(b)} Waterfall plot showing the instantaneous frequency of the demodulated signal when the laser current is modulated at 0.1~Hz showing that the laser behaves as a current-controlled oscillator. \textbf{(c)} Original and received audio signal (see Supplementary Material for the received audio file: ``Volare" by Dean Martin).}
\label{fig3_radio}
\end{figure}

\begin{figure*}[t]
\centering
\includegraphics[width=.9\textwidth]{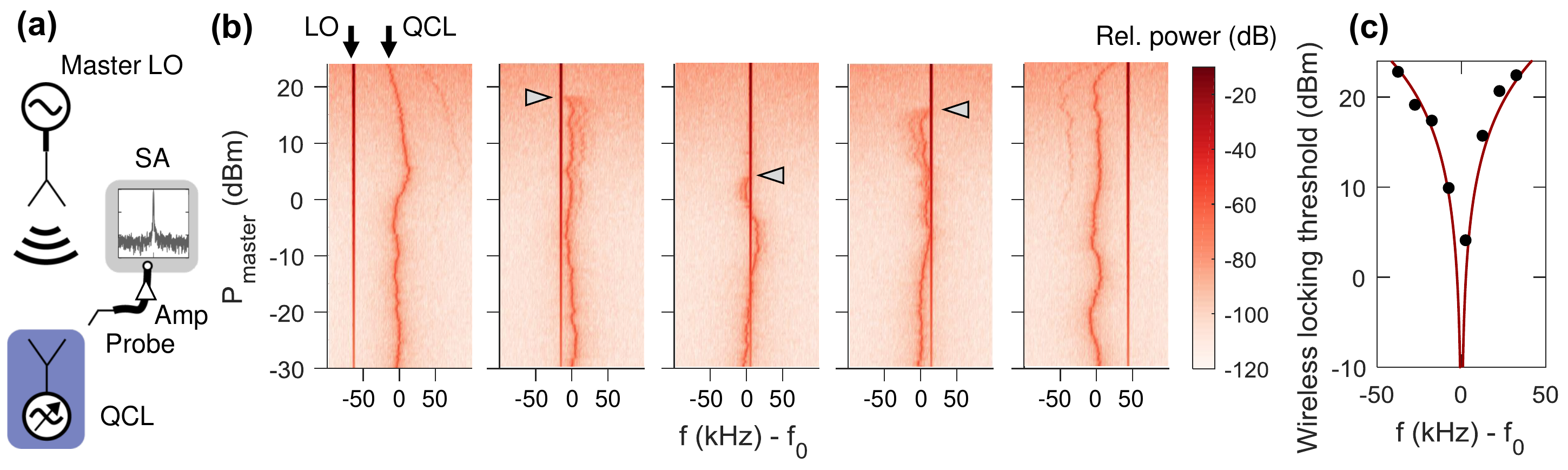}
\caption{\textbf{(a)} Schematic of the set-up to injection lock the QCL beat note to a local oscillator (LO) through free space. A probe placed near the antenna of the QCL monitors the changes in the laser beat note induced by the LO. \textbf{(b)} Shift of the QCL beat note spectrum as the power of the LO is swept between -30 and 24~dBm. The LO power threshold at which locking occurs is marked by arrowheads. Frequencies are given with respect to the beat note frequency of the free-running QCL ($f_0 = 5.501$~GHz). Five exemplary cases are shown. \textbf{(c)} Measured LO power corresponding to the threshold for wireless locking of the QCL beat note to the LO for different LO frequencies (circles). Also shown is the fit of the theoretical quadratic dependence (continuous line).}
\label{fig4_locking}
\end{figure*}

Next, we provide a proof-of-concept of wireless communication using the laser radio transmitter. The experimental set-up is schematized in Fig.~\ref{fig3_radio}a. The laser current is modulated by an audio analog signal which in turn modulates the frequency of the laser beat note allowing to encode the baseband information onto the 5.5~GHz carrier wave. The radio signal is received by a horn antenna at a distance of 0.9~m, filtered by a receiving filter (1.9-5.5~GHz bandwidth), amplified and then downconverted to 1.5~GHz by mixing with a local oscillator (7.0~GHz) to fit into the bandwidth of a software-defined radio (SDR) used for demodulation. The physical process underlying the current-driven beat note modulation is the following: the current modulation $\Delta I$ induces a thermal variation of the QCL active region, thus changing the group refractive index $n_g$ of the cavity. This in turn changes the intermodal spacing of the comb and the beat note frequency. In Fig.~\ref{fig3_radio}b we present a waterfall plot of the signal demodulated by the SDR when the laser current is modulated at $f_{mod}=0.1$~Hz with a relative current modulation amplitude of $\Delta I/I=0.2\%$. The beat note exhibits a nearly constant power, and its instantaneous frequency is modulated with a period of 10~s defined by $f_{mod}$, spanning a range of frequency deviation of 120~kHz determined by $\Delta I$. In essence the QCL behaves as a current-controlled oscillator which can generate a frequency modulated (FM) signal. This scheme is used for radio transmission of an audio track, which can be correctly retrieved after demodulation (Fig.~\ref{fig3_radio}c; see Supplementary Material for the audio file). Unwanted slow thermal fluctuations of the laser cavity, which persist despite the use of a temperature controller, induce a jittering of the beat note appearing as a slow modulation of the baseline of the received audio signal. The effect of these fluctuations is to add a noise contribution below 10~Hz (Supplementary Sec.~5), which lies outside of the audio frequency range (20-20'000~Hz). The high-frequency background noise that is audible in the track is due to the level of the noise floor. The signal-to-noise ratio could be improved by increasing the range of frequency deviation of the modulated beat note, though this was limited in the present work by the bandwidth of the software demodulator. While the present demonstration deals with a low-frequency audio modulation signal, QCL modulation of several tens of gigahertz has been demonstrated using microstrip and coplanar waveguide geometries~\cite{Maineult2010,Gellie:10,Calvar2013,Hinkov:16}. These studies investigated the high frequency modulation of the optical field in single-mode QCLs and can aid in understanding the effect of multimode QCL modulation on the generated beat note. In particular, at gigahertz modulation frequencies, we expect that the laser beat note will be predominantly amplitude modulated by plasma effects, as the cut-off frequency of thermal effects responsible for frequency modulation lies in the megahertz range~\cite{Hangauer:14}.

Due to the presence of the antenna, the laser is also sensitive to wireless radio frequency signals. Here we show that the laser beat note can be wirelessly injection locked to an external microwave reference. A schematic of the set-up is shown in Fig.~\ref{fig4_locking}a. A local oscillator (LO), tunable in power and frequency, is connected to a horn antenna directed at the QCL chip. A microwave probe is placed in proximity of the QCL antenna to monitor the changes in the laser beat note induced by the LO, whose power is swept between -30 and 24~dBm for a set of different frequencies. Examples of the behavior exhibited by the QCL beat note upon wireless microwave injection are shown in Fig.~\ref{fig4_locking}b. In the lower range of powers of the LO the QCL behaves as free-running, its beat note is unlocked and shows small frequency oscillations around $f_0 = 5.501$~GHz due to thermal fluctuations. When the LO power overcomes a threshold dependent on the detuning from $f_0$ the QCL beat note locks to the external oscillator and stops jittering. This phenomenon is preceded by the appearance of a weaker sideband next to the QCL beat note, as typically observed in wired injection-locking experiments~\cite{Gellie:10}. The increase of the power threshold with the detuning from $f_0$ follows a quadratic dependence, as expected from injection-locking theory~\cite{Siegman1986,AdlerOct.}. For the maximum explored power of the LO (24~dBm) the wireless locking range is about 40~kHz. We note that an enhancement of the microwave emission from the laser based on a buried heterostructure design as discussed above, will also improve the coupling of external radio signals into the system, thus lowering the wireless locking threshold curve. The demonstration of wireless injection locking of the laser beat note shows the possibility of remote control of laser frequency combs and may open new applications in the field such as wireless synchronization of multiple comb generators to a single reference oscillator, without the need of integrating complex interconnected microwave architectures.

This work is a proof of concept demonstration of a QCL frequency comb utilized as a wireless radio transmitter. Thanks to the recently discovered harmonic comb operation of QCLs~\cite{Kazakov2017,Piccardo:18} --- where the intermodal comb spacing lies in the hundreds of gigahertz range due to longitudinal mode skipping --- the frequency range of radio transmission of the system can be potentially extended to sub-terahertz carriers. Moreover the frequency of such carriers holds promise of broad tunability, since the intermodal spacing of a harmonic frequency comb may be varied from few 100 GHz up to over 1 THz in a single device, as it was recently experimentally demonstrated by optical injection of an external seed~\cite{Piccardo2018tunable}. In the harmonic regime the internal dynamic grating will exhibit a number of spatial cycles corresponding to the number of skipped longitudinal modes in the frequency comb~\cite{Piccardo:18}. Adapting the antenna design demonstrated here to match the spatial periodicity of this higher-order grating will provide for extraction of a substantial fraction of the available radio frequency power generated in the harmonic state. For instance, travelling- and standing-wave antenna designs, which have already been proven successful in cryogenically-cooled terahertz QCL systems~\cite{Tavallaee2011,Masini2017}, or surface grating outcouplers, used in difference-frequency-generation terahertz QCLs~\cite{Pflugl2008,Kim:18}, could be used to efficiently radiate the generated sub-terahertz signal. Besides increasing the carrier wave frequency, the harmonic state offers a spectrum constituted by few, powerful optical modes. In principle, this spectral distribution of the optical power presents the net advantage of a much more efficient radio frequency generation as compared to QCLs operating in the fundamental comb regime. Numerical simulations suggest that a QCL in the harmonic state can generate a beat note in the 100~GHz range with a 37~dB power enhancement with respect to that produced at 5.5~GHz by a QCL operating as a fundamental frequency comb in similar current and optical power conditions (Supplementary Sec.~4).

QCL radio frequency sources benefit from good impedance-matching with extraction elements such as antennas and waveguides, thanks to the low impedance of their active region. This is a clear advantage with respect to existing terahertz photomixers, which suffer from high impedance in the order of tens of kiloohm, thus losing several orders of magnitude in power efficiency. Thanks to the frequency comb nature of the light beating inside the cavity, this new radio frequency source can generate tones of high spectral purity, leading to a very narrow (kHz or sub-kHz) linewidth. Another attractive feature of the LRT is that the carrier frequency may be controlled by the laser current, allowing in principle to phase-lock it to a reference microwave source using frequency division and thus stabilize it with high accuracy. Ultimately, the system introduced here will benefit from unprecedented compactness, as compared to existing composite terahertz wireless communication systems~\cite{Nagatsuma2016}, unifying in a single device the capability of room-temperature generation, modulation and emission of sub-terahertz waves, and may find applications in fields ranging from telecommunications and spectroscopy, to radioastronomy and quantum optics.

This work was supported by the National Science Foundation (Grants No. ECCS-1614631, 1541959, DGE1144152), Assistant Secretary of Defense for Research and Engineering (Air Force Contract FA8721-05-C-0002 and/or FA8702-15D-0001), Swiss National Science Foundation (177836), Austrian Science Fund (P28914−N27). We thank D. Kazakov for discussions that motivated this demonstration and for a careful reading of the manuscript, and A. Amirzhan for useful discussions.


\bibliography{wirelessbib}

\begin{thebibliography}{34}%
\makeatletter
\providecommand \@ifxundefined [1]{%
 \@ifx{#1\undefined}
}%
\providecommand \@ifnum [1]{%
 \ifnum #1\expandafter \@firstoftwo
 \else \expandafter \@secondoftwo
 \fi
}%
\providecommand \@ifx [1]{%
 \ifx #1\expandafter \@firstoftwo
 \else \expandafter \@secondoftwo
 \fi
}%
\providecommand \natexlab [1]{#1}%
\providecommand \enquote  [1]{``#1''}%
\providecommand \bibnamefont  [1]{#1}%
\providecommand \bibfnamefont [1]{#1}%
\providecommand \citenamefont [1]{#1}%
\providecommand \href@noop [0]{\@secondoftwo}%
\providecommand \href [0]{\begingroup \@sanitize@url \@href}%
\providecommand \@href[1]{\@@startlink{#1}\@@href}%
\providecommand \@@href[1]{\endgroup#1\@@endlink}%
\providecommand \@sanitize@url [0]{\catcode `\\12\catcode `\$12\catcode
  `\&12\catcode `\#12\catcode `\^12\catcode `\_12\catcode `\%12\relax}%
\providecommand \@@startlink[1]{}%
\providecommand \@@endlink[0]{}%
\providecommand \url  [0]{\begingroup\@sanitize@url \@url }%
\providecommand \@url [1]{\endgroup\@href {#1}{\urlprefix }}%
\providecommand \urlprefix  [0]{URL }%
\providecommand \Eprint [0]{\href }%
\providecommand \doibase [0]{http://dx.doi.org/}%
\providecommand \selectlanguage [0]{\@gobble}%
\providecommand \bibinfo  [0]{\@secondoftwo}%
\providecommand \bibfield  [0]{\@secondoftwo}%
\providecommand \translation [1]{[#1]}%
\providecommand \BibitemOpen [0]{}%
\providecommand \bibitemStop [0]{}%
\providecommand \bibitemNoStop [0]{.\EOS\space}%
\providecommand \EOS [0]{\spacefactor3000\relax}%
\providecommand \BibitemShut  [1]{\csname bibitem#1\endcsname}%
\let\auto@bib@innerbib\@empty
\bibitem [{\citenamefont {Fortier}\ \emph {et~al.}(2011)\citenamefont
  {Fortier}, \citenamefont {Kirchner}, \citenamefont {Quinlan}, \citenamefont
  {Taylor}, \citenamefont {Bergquist}, \citenamefont {Rosenband}, \citenamefont
  {Lemke}, \citenamefont {Ludlow}, \citenamefont {Jiang}, \citenamefont
  {Oates},\ and\ \citenamefont {Diddams}}]{Fortier2011}%
  \BibitemOpen
  \bibfield  {author} {\bibinfo {author} {\bibfnamefont {T.~M.}\ \bibnamefont
  {Fortier}}, \bibinfo {author} {\bibfnamefont {M.~S.}\ \bibnamefont
  {Kirchner}}, \bibinfo {author} {\bibfnamefont {F.}~\bibnamefont {Quinlan}},
  \bibinfo {author} {\bibfnamefont {J.}~\bibnamefont {Taylor}}, \bibinfo
  {author} {\bibfnamefont {J.~C.}\ \bibnamefont {Bergquist}}, \bibinfo {author}
  {\bibfnamefont {T.}~\bibnamefont {Rosenband}}, \bibinfo {author}
  {\bibfnamefont {N.}~\bibnamefont {Lemke}}, \bibinfo {author} {\bibfnamefont
  {A.}~\bibnamefont {Ludlow}}, \bibinfo {author} {\bibfnamefont
  {Y.}~\bibnamefont {Jiang}}, \bibinfo {author} {\bibfnamefont {C.~W.}\
  \bibnamefont {Oates}}, \ and\ \bibinfo {author} {\bibfnamefont {S.~A.}\
  \bibnamefont {Diddams}},\ }\href {http://dx.doi.org/10.1038/nphoton.2011.121}
  {\bibfield  {journal} {\bibinfo  {journal} {Nat. Photon.}\ }\textbf {\bibinfo
  {volume} {5}},\ \bibinfo {pages} {425} (\bibinfo {year} {2011})}\BibitemShut
  {NoStop}%
\bibitem [{\citenamefont {Li}\ \emph {et~al.}(2014)\citenamefont {Li},
  \citenamefont {Yi}, \citenamefont {Lee}, \citenamefont {Diddams},\ and\
  \citenamefont {Vahala}}]{Li309}%
  \BibitemOpen
  \bibfield  {author} {\bibinfo {author} {\bibfnamefont {J.}~\bibnamefont
  {Li}}, \bibinfo {author} {\bibfnamefont {X.}~\bibnamefont {Yi}}, \bibinfo
  {author} {\bibfnamefont {H.}~\bibnamefont {Lee}}, \bibinfo {author}
  {\bibfnamefont {S.~A.}\ \bibnamefont {Diddams}}, \ and\ \bibinfo {author}
  {\bibfnamefont {K.~J.}\ \bibnamefont {Vahala}},\ }\href {\doibase
  10.1126/science.1252909} {\bibfield  {journal} {\bibinfo  {journal}
  {Science}\ }\textbf {\bibinfo {volume} {345}},\ \bibinfo {pages} {309}
  (\bibinfo {year} {2014})}\BibitemShut {NoStop}%
\bibitem [{\citenamefont {Maleki}(2011)}]{Maleki2011}%
  \BibitemOpen
  \bibfield  {author} {\bibinfo {author} {\bibfnamefont {L.}~\bibnamefont
  {Maleki}},\ }\href {http://dx.doi.org/10.1038/nphoton.2011.293} {\bibfield
  {journal} {\bibinfo  {journal} {Nat. Photon.}\ }\textbf {\bibinfo {volume}
  {5}},\ \bibinfo {pages} {728} (\bibinfo {year} {2011})}\BibitemShut {NoStop}%
\bibitem [{\citenamefont {Li}\ \emph {et~al.}(2013)\citenamefont {Li},
  \citenamefont {Lee},\ and\ \citenamefont {Vahala}}]{Li2013}%
  \BibitemOpen
  \bibfield  {author} {\bibinfo {author} {\bibfnamefont {J.}~\bibnamefont
  {Li}}, \bibinfo {author} {\bibfnamefont {H.}~\bibnamefont {Lee}}, \ and\
  \bibinfo {author} {\bibfnamefont {K.~J.}\ \bibnamefont {Vahala}},\ }\href
  {http://dx.doi.org/10.1038/ncomms3097} {\bibfield  {journal} {\bibinfo
  {journal} {Nat. Commun.}\ }\textbf {\bibinfo {volume} {4}},\ \bibinfo {pages}
  {2097} (\bibinfo {year} {2013})}\BibitemShut {NoStop}%
\bibitem [{\citenamefont {Tani}\ \emph {et~al.}(2005)\citenamefont {Tani},
  \citenamefont {Morikawa}, \citenamefont {Matsuura},\ and\ \citenamefont
  {Hangyo}}]{Tani2005}%
  \BibitemOpen
  \bibfield  {author} {\bibinfo {author} {\bibfnamefont {M.}~\bibnamefont
  {Tani}}, \bibinfo {author} {\bibfnamefont {O.}~\bibnamefont {Morikawa}},
  \bibinfo {author} {\bibfnamefont {S.}~\bibnamefont {Matsuura}}, \ and\
  \bibinfo {author} {\bibfnamefont {M.}~\bibnamefont {Hangyo}},\ }\href
  {http://stacks.iop.org/0268-1242/20/i=7/a=005} {\bibfield  {journal}
  {\bibinfo  {journal} {Semicond. Sci. Tech.}\ }\textbf {\bibinfo {volume}
  {20}},\ \bibinfo {pages} {151} (\bibinfo {year} {2005})}\BibitemShut
  {NoStop}%
\bibitem [{\citenamefont {Kim}\ and\ \citenamefont
  {K\"{a}rtner}(2010)}]{Kim:10}%
  \BibitemOpen
  \bibfield  {author} {\bibinfo {author} {\bibfnamefont {J.}~\bibnamefont
  {Kim}}\ and\ \bibinfo {author} {\bibfnamefont {F.~X.}\ \bibnamefont
  {K\"{a}rtner}},\ }\href {\doibase 10.1364/OL.35.002022} {\bibfield  {journal}
  {\bibinfo  {journal} {Opt. Lett.}\ }\textbf {\bibinfo {volume} {35}},\
  \bibinfo {pages} {2022} (\bibinfo {year} {2010})}\BibitemShut {NoStop}%
\bibitem [{\citenamefont {Rubiola}\ and\ \citenamefont
  {Santarelli}(2013)}]{Rubiola2013}%
  \BibitemOpen
  \bibfield  {author} {\bibinfo {author} {\bibfnamefont {E.}~\bibnamefont
  {Rubiola}}\ and\ \bibinfo {author} {\bibfnamefont {G.}~\bibnamefont
  {Santarelli}},\ }\href@noop {} {\bibfield  {journal} {\bibinfo  {journal}
  {Nat. Photon.}\ }\textbf {\bibinfo {volume} {7}},\ \bibinfo {pages} {269}
  (\bibinfo {year} {2013})}\BibitemShut {NoStop}%
\bibitem [{\citenamefont {Quinlan}\ \emph {et~al.}(2013)\citenamefont
  {Quinlan}, \citenamefont {Fortier}, \citenamefont {Jiang}, \citenamefont
  {Hati}, \citenamefont {Nelson}, \citenamefont {Fu}, \citenamefont
  {Campbell},\ and\ \citenamefont {Diddams}}]{Quinlan2013}%
  \BibitemOpen
  \bibfield  {author} {\bibinfo {author} {\bibfnamefont {F.}~\bibnamefont
  {Quinlan}}, \bibinfo {author} {\bibfnamefont {T.~M.}\ \bibnamefont
  {Fortier}}, \bibinfo {author} {\bibfnamefont {H.}~\bibnamefont {Jiang}},
  \bibinfo {author} {\bibfnamefont {A.}~\bibnamefont {Hati}}, \bibinfo {author}
  {\bibfnamefont {C.}~\bibnamefont {Nelson}}, \bibinfo {author} {\bibfnamefont
  {Y.}~\bibnamefont {Fu}}, \bibinfo {author} {\bibfnamefont {J.~C.}\
  \bibnamefont {Campbell}}, \ and\ \bibinfo {author} {\bibfnamefont {S.~A.}\
  \bibnamefont {Diddams}},\ }\href {http://dx.doi.org/10.1038/nphoton.2013.33}
  {\bibfield  {journal} {\bibinfo  {journal} {Nat. Photon.}\ }\textbf {\bibinfo
  {volume} {7}},\ \bibinfo {pages} {290} (\bibinfo {year} {2013})}\BibitemShut
  {NoStop}%
\bibitem [{\citenamefont {Udem}\ \emph {et~al.}(2002)\citenamefont {Udem},
  \citenamefont {Holzwarth},\ and\ \citenamefont {H{\"{a}}nsch}}]{Udem2002}%
  \BibitemOpen
  \bibfield  {author} {\bibinfo {author} {\bibfnamefont {T.}~\bibnamefont
  {Udem}}, \bibinfo {author} {\bibfnamefont {R.}~\bibnamefont {Holzwarth}}, \
  and\ \bibinfo {author} {\bibfnamefont {T.~W.}\ \bibnamefont {H{\"{a}}nsch}},\
  }\href {\doibase 10.1038/416233a} {\bibfield  {journal} {\bibinfo  {journal}
  {Nature}\ }\textbf {\bibinfo {volume} {416}},\ \bibinfo {pages} {233}
  (\bibinfo {year} {2002})}\BibitemShut {NoStop}%
\bibitem [{\citenamefont {H{\"{a}}nsch}(2006)}]{Hansch2006}%
  \BibitemOpen
  \bibfield  {author} {\bibinfo {author} {\bibfnamefont {T.~W.}\ \bibnamefont
  {H{\"{a}}nsch}},\ }\href {\doibase 10.1103/RevModPhys.78.1297} {\bibfield
  {journal} {\bibinfo  {journal} {Rev. Mod. Phys.}\ }\textbf {\bibinfo {volume}
  {78}},\ \bibinfo {pages} {1297} (\bibinfo {year} {2006})}\BibitemShut
  {NoStop}%
\bibitem [{\citenamefont {Hugi}\ \emph {et~al.}(2012)\citenamefont {Hugi},
  \citenamefont {Villares}, \citenamefont {Blaser}, \citenamefont {Liu},\ and\
  \citenamefont {Faist}}]{Hugi2012}%
  \BibitemOpen
  \bibfield  {author} {\bibinfo {author} {\bibfnamefont {A.}~\bibnamefont
  {Hugi}}, \bibinfo {author} {\bibfnamefont {G.}~\bibnamefont {Villares}},
  \bibinfo {author} {\bibfnamefont {S.}~\bibnamefont {Blaser}}, \bibinfo
  {author} {\bibfnamefont {H.~C.}\ \bibnamefont {Liu}}, \ and\ \bibinfo
  {author} {\bibfnamefont {J.}~\bibnamefont {Faist}},\ }\href {\doibase
  10.1038/nature11620} {\bibfield  {journal} {\bibinfo  {journal} {Nature}\
  }\textbf {\bibinfo {volume} {492}},\ \bibinfo {pages} {229} (\bibinfo {year}
  {2012})}\BibitemShut {NoStop}%
\bibitem [{\citenamefont {Villares}\ \emph {et~al.}(2014)\citenamefont
  {Villares}, \citenamefont {Hugi}, \citenamefont {Blaser},\ and\ \citenamefont
  {Faist}}]{Villares2014}%
  \BibitemOpen
  \bibfield  {author} {\bibinfo {author} {\bibfnamefont {G.}~\bibnamefont
  {Villares}}, \bibinfo {author} {\bibfnamefont {A.}~\bibnamefont {Hugi}},
  \bibinfo {author} {\bibfnamefont {S.}~\bibnamefont {Blaser}}, \ and\ \bibinfo
  {author} {\bibfnamefont {J.}~\bibnamefont {Faist}},\ }\href {\doibase
  10.1038/ncomms6192} {\bibfield  {journal} {\bibinfo  {journal} {Nat.
  Commun.}\ }\textbf {\bibinfo {volume} {5}},\ \bibinfo {pages} {5192}
  (\bibinfo {year} {2014})}\BibitemShut {NoStop}%
\bibitem [{\citenamefont {Burghoff}\ \emph {et~al.}(2014)\citenamefont
  {Burghoff}, \citenamefont {Kao}, \citenamefont {Han}, \citenamefont {Chan},
  \citenamefont {Cai}, \citenamefont {Yang}, \citenamefont {Hayton},
  \citenamefont {Gao}, \citenamefont {Reno},\ and\ \citenamefont
  {Hu}}]{Burghoff2014}%
  \BibitemOpen
  \bibfield  {author} {\bibinfo {author} {\bibfnamefont {D.}~\bibnamefont
  {Burghoff}}, \bibinfo {author} {\bibfnamefont {T.-Y.}\ \bibnamefont {Kao}},
  \bibinfo {author} {\bibfnamefont {N.}~\bibnamefont {Han}}, \bibinfo {author}
  {\bibfnamefont {C.~W.~I.}\ \bibnamefont {Chan}}, \bibinfo {author}
  {\bibfnamefont {X.}~\bibnamefont {Cai}}, \bibinfo {author} {\bibfnamefont
  {Y.}~\bibnamefont {Yang}}, \bibinfo {author} {\bibfnamefont {D.~J.}\
  \bibnamefont {Hayton}}, \bibinfo {author} {\bibfnamefont {J.-R.}\
  \bibnamefont {Gao}}, \bibinfo {author} {\bibfnamefont {J.~L.}\ \bibnamefont
  {Reno}}, \ and\ \bibinfo {author} {\bibfnamefont {Q.}~\bibnamefont {Hu}},\
  }\href {http://dx.doi.org/10.1038/nphoton.2014.85} {\bibfield  {journal}
  {\bibinfo  {journal} {Nat. Photon.}\ }\textbf {\bibinfo {volume} {8}},\
  \bibinfo {pages} {462} (\bibinfo {year} {2014})}\BibitemShut {NoStop}%
\bibitem [{\citenamefont {Lu}\ \emph {et~al.}(2015)\citenamefont {Lu},
  \citenamefont {Razeghi}, \citenamefont {Slivken}, \citenamefont
  {Bandyopadhyay}, \citenamefont {Bai}, \citenamefont {Zhou}, \citenamefont
  {Chen}, \citenamefont {Heydari}, \citenamefont {Haddadi}, \citenamefont
  {McClintock}, \citenamefont {Amanti},\ and\ \citenamefont
  {Sirtori}}]{Lu2015}%
  \BibitemOpen
  \bibfield  {author} {\bibinfo {author} {\bibfnamefont {Q.~Y.}\ \bibnamefont
  {Lu}}, \bibinfo {author} {\bibfnamefont {M.}~\bibnamefont {Razeghi}},
  \bibinfo {author} {\bibfnamefont {S.}~\bibnamefont {Slivken}}, \bibinfo
  {author} {\bibfnamefont {N.}~\bibnamefont {Bandyopadhyay}}, \bibinfo {author}
  {\bibfnamefont {Y.}~\bibnamefont {Bai}}, \bibinfo {author} {\bibfnamefont
  {W.~J.}\ \bibnamefont {Zhou}}, \bibinfo {author} {\bibfnamefont
  {M.}~\bibnamefont {Chen}}, \bibinfo {author} {\bibfnamefont {D.}~\bibnamefont
  {Heydari}}, \bibinfo {author} {\bibfnamefont {A.}~\bibnamefont {Haddadi}},
  \bibinfo {author} {\bibfnamefont {R.}~\bibnamefont {McClintock}}, \bibinfo
  {author} {\bibfnamefont {M.}~\bibnamefont {Amanti}}, \ and\ \bibinfo {author}
  {\bibfnamefont {C.}~\bibnamefont {Sirtori}},\ }\href {\doibase
  10.1063/1.4907646} {\bibfield  {journal} {\bibinfo  {journal} {Appl. Phys.
  Lett.}\ }\textbf {\bibinfo {volume} {106}},\ \bibinfo {pages} {51105}
  (\bibinfo {year} {2015})}\BibitemShut {NoStop}%
\bibitem [{\citenamefont {Cappelli}\ \emph {et~al.}(2016)\citenamefont
  {Cappelli}, \citenamefont {Campo}, \citenamefont {Galli}, \citenamefont
  {Giusfredi}, \citenamefont {Bartalini}, \citenamefont {Mazzotti},
  \citenamefont {Cancio}, \citenamefont {Borri}, \citenamefont {Hinkov},
  \citenamefont {Faist},\ and\ \citenamefont {De~Natale}}]{Cappelli2016}%
  \BibitemOpen
  \bibfield  {author} {\bibinfo {author} {\bibfnamefont {F.}~\bibnamefont
  {Cappelli}}, \bibinfo {author} {\bibfnamefont {G.}~\bibnamefont {Campo}},
  \bibinfo {author} {\bibfnamefont {I.}~\bibnamefont {Galli}}, \bibinfo
  {author} {\bibfnamefont {G.}~\bibnamefont {Giusfredi}}, \bibinfo {author}
  {\bibfnamefont {S.}~\bibnamefont {Bartalini}}, \bibinfo {author}
  {\bibfnamefont {D.}~\bibnamefont {Mazzotti}}, \bibinfo {author}
  {\bibfnamefont {P.}~\bibnamefont {Cancio}}, \bibinfo {author} {\bibfnamefont
  {S.}~\bibnamefont {Borri}}, \bibinfo {author} {\bibfnamefont
  {B.}~\bibnamefont {Hinkov}}, \bibinfo {author} {\bibfnamefont
  {J.}~\bibnamefont {Faist}}, \ and\ \bibinfo {author} {\bibfnamefont
  {P.}~\bibnamefont {De~Natale}},\ }\href@noop {} {\bibfield  {journal}
  {\bibinfo  {journal} {Laser \& Photonics Reviews}\ }\textbf {\bibinfo
  {volume} {10}},\ \bibinfo {pages} {623} (\bibinfo {year} {2016})}\BibitemShut
  {NoStop}%
\bibitem [{\citenamefont {Piccardo}\ \emph
  {et~al.}(2018{\natexlab{a}})\citenamefont {Piccardo}, \citenamefont
  {Kazakov}, \citenamefont {Rubin}, \citenamefont {Chevalier}, \citenamefont
  {Wang}, \citenamefont {Xie}, \citenamefont {Lascola}, \citenamefont
  {Belyanin},\ and\ \citenamefont {Capasso}}]{Piccardo:18}%
  \BibitemOpen
  \bibfield  {author} {\bibinfo {author} {\bibfnamefont {M.}~\bibnamefont
  {Piccardo}}, \bibinfo {author} {\bibfnamefont {D.}~\bibnamefont {Kazakov}},
  \bibinfo {author} {\bibfnamefont {N.~A.}\ \bibnamefont {Rubin}}, \bibinfo
  {author} {\bibfnamefont {P.}~\bibnamefont {Chevalier}}, \bibinfo {author}
  {\bibfnamefont {Y.}~\bibnamefont {Wang}}, \bibinfo {author} {\bibfnamefont
  {F.}~\bibnamefont {Xie}}, \bibinfo {author} {\bibfnamefont {K.}~\bibnamefont
  {Lascola}}, \bibinfo {author} {\bibfnamefont {A.}~\bibnamefont {Belyanin}}, \
  and\ \bibinfo {author} {\bibfnamefont {F.}~\bibnamefont {Capasso}},\ }\href
  {\doibase 10.1364/OPTICA.5.000475} {\bibfield  {journal} {\bibinfo  {journal}
  {Optica}\ }\textbf {\bibinfo {volume} {5}},\ \bibinfo {pages} {475} (\bibinfo
  {year} {2018}{\natexlab{a}})}\BibitemShut {NoStop}%
\bibitem [{\citenamefont {Choi}\ \emph {et~al.}(2008)\citenamefont {Choi},
  \citenamefont {Diehl}, \citenamefont {Wu}, \citenamefont {Giovannini},
  \citenamefont {Faist}, \citenamefont {Capasso},\ and\ \citenamefont
  {Norris}}]{Choi2008PRL}%
  \BibitemOpen
  \bibfield  {author} {\bibinfo {author} {\bibfnamefont {H.}~\bibnamefont
  {Choi}}, \bibinfo {author} {\bibfnamefont {L.}~\bibnamefont {Diehl}},
  \bibinfo {author} {\bibfnamefont {Z.-K.}\ \bibnamefont {Wu}}, \bibinfo
  {author} {\bibfnamefont {M.}~\bibnamefont {Giovannini}}, \bibinfo {author}
  {\bibfnamefont {J.}~\bibnamefont {Faist}}, \bibinfo {author} {\bibfnamefont
  {F.}~\bibnamefont {Capasso}}, \ and\ \bibinfo {author} {\bibfnamefont
  {T.~B.}\ \bibnamefont {Norris}},\ }\href {\doibase
  10.1103/PhysRevLett.100.167401} {\bibfield  {journal} {\bibinfo  {journal}
  {Phys. Rev. Lett.}\ }\textbf {\bibinfo {volume} {100}},\ \bibinfo {pages}
  {167401} (\bibinfo {year} {2008})}\BibitemShut {NoStop}%
\bibitem [{\citenamefont {Kazakov}\ \emph {et~al.}(2017)\citenamefont
  {Kazakov}, \citenamefont {Piccardo}, \citenamefont {Chevalier}, \citenamefont
  {Mansuripur}, \citenamefont {Wang}, \citenamefont {Xie}, \citenamefont
  {en~Zah}, \citenamefont {Lascola}, \citenamefont {Belyanin},\ and\
  \citenamefont {Capasso}}]{Kazakov2017}%
  \BibitemOpen
  \bibfield  {author} {\bibinfo {author} {\bibfnamefont {D.}~\bibnamefont
  {Kazakov}}, \bibinfo {author} {\bibfnamefont {M.}~\bibnamefont {Piccardo}},
  \bibinfo {author} {\bibfnamefont {P.}~\bibnamefont {Chevalier}}, \bibinfo
  {author} {\bibfnamefont {T.~S.}\ \bibnamefont {Mansuripur}}, \bibinfo
  {author} {\bibfnamefont {Y.}~\bibnamefont {Wang}}, \bibinfo {author}
  {\bibfnamefont {F.}~\bibnamefont {Xie}}, \bibinfo {author} {\bibfnamefont
  {C.}~\bibnamefont {en~Zah}}, \bibinfo {author} {\bibfnamefont
  {K.}~\bibnamefont {Lascola}}, \bibinfo {author} {\bibfnamefont
  {A.}~\bibnamefont {Belyanin}}, \ and\ \bibinfo {author} {\bibfnamefont
  {F.}~\bibnamefont {Capasso}},\ }\href@noop {} {\bibfield  {journal} {\bibinfo
   {journal} {Nat. Photon.}\ }\textbf {\bibinfo {volume} {11}},\ \bibinfo
  {pages} {789} (\bibinfo {year} {2017})}\BibitemShut {NoStop}%
\bibitem [{\citenamefont {Nagatsuma}\ \emph {et~al.}(2016)\citenamefont
  {Nagatsuma}, \citenamefont {Ducournau},\ and\ \citenamefont
  {Renaud}}]{Nagatsuma2016}%
  \BibitemOpen
  \bibfield  {author} {\bibinfo {author} {\bibfnamefont {T.}~\bibnamefont
  {Nagatsuma}}, \bibinfo {author} {\bibfnamefont {G.}~\bibnamefont
  {Ducournau}}, \ and\ \bibinfo {author} {\bibfnamefont {C.~C.}\ \bibnamefont
  {Renaud}},\ }\href
  {https://www.nature.com/nphoton/journal/v10/n6/pdf/nphoton.2016.65.pdf}
  {\bibfield  {journal} {\bibinfo  {journal} {Nat. Photon.}\ }\textbf {\bibinfo
  {volume} {10}},\ \bibinfo {pages} {371} (\bibinfo {year} {2016})}\BibitemShut
  {NoStop}%
\bibitem [{\citenamefont {Akyildiz}\ \emph {et~al.}(2014)\citenamefont
  {Akyildiz}, \citenamefont {Jornet},\ and\ \citenamefont
  {Han}}]{Akyildiz2014}%
  \BibitemOpen
  \bibfield  {author} {\bibinfo {author} {\bibfnamefont {I.~F.}\ \bibnamefont
  {Akyildiz}}, \bibinfo {author} {\bibfnamefont {J.~M.}\ \bibnamefont
  {Jornet}}, \ and\ \bibinfo {author} {\bibfnamefont {C.}~\bibnamefont {Han}},\
  }\href {\doibase https://doi.org/10.1016/j.phycom.2014.01.006} {\bibfield
  {journal} {\bibinfo  {journal} {Physical Communication}\ }\textbf {\bibinfo
  {volume} {12}},\ \bibinfo {pages} {16} (\bibinfo {year} {2014})}\BibitemShut
  {NoStop}%
\bibitem [{\citenamefont {Kleine-Ostmann}\ and\ \citenamefont
  {Nagatsuma}(2011)}]{Kleine-Ostmann2011}%
  \BibitemOpen
  \bibfield  {author} {\bibinfo {author} {\bibfnamefont {T.}~\bibnamefont
  {Kleine-Ostmann}}\ and\ \bibinfo {author} {\bibfnamefont {T.}~\bibnamefont
  {Nagatsuma}},\ }\href {\doibase 10.1007/s10762-010-9758-1} {\bibfield
  {journal} {\bibinfo  {journal} {J. Infrared Millim. Terahertz Waves}\
  }\textbf {\bibinfo {volume} {32}},\ \bibinfo {pages} {143} (\bibinfo {year}
  {2011})}\BibitemShut {NoStop}%
\bibitem [{\citenamefont {Nida}\ \emph {et~al.}(2017)\citenamefont {Nida},
  \citenamefont {Hinkov}, \citenamefont {Gini},\ and\ \citenamefont
  {Faist}}]{Nida2017}%
  \BibitemOpen
  \bibfield  {author} {\bibinfo {author} {\bibfnamefont {S.}~\bibnamefont
  {Nida}}, \bibinfo {author} {\bibfnamefont {B.}~\bibnamefont {Hinkov}},
  \bibinfo {author} {\bibfnamefont {E.}~\bibnamefont {Gini}}, \ and\ \bibinfo
  {author} {\bibfnamefont {J.}~\bibnamefont {Faist}},\ }\href {\doibase
  10.1063/1.4977243} {\bibfield  {journal} {\bibinfo  {journal} {J. Appl.
  Phys.}\ }\textbf {\bibinfo {volume} {121}},\ \bibinfo {pages} {094502}
  (\bibinfo {year} {2017})}\BibitemShut {NoStop}%
\bibitem [{\citenamefont {Maineult}\ \emph {et~al.}(2010)\citenamefont
  {Maineult}, \citenamefont {Ding}, \citenamefont {Gellie}, \citenamefont
  {Filloux}, \citenamefont {Sirtori}, \citenamefont {Barbieri}, \citenamefont
  {Akalin}, \citenamefont {Lampin}, \citenamefont {Sagnes}, \citenamefont
  {Beere},\ and\ \citenamefont {Ritchie}}]{Maineult2010}%
  \BibitemOpen
  \bibfield  {author} {\bibinfo {author} {\bibfnamefont {W.}~\bibnamefont
  {Maineult}}, \bibinfo {author} {\bibfnamefont {L.}~\bibnamefont {Ding}},
  \bibinfo {author} {\bibfnamefont {P.}~\bibnamefont {Gellie}}, \bibinfo
  {author} {\bibfnamefont {P.}~\bibnamefont {Filloux}}, \bibinfo {author}
  {\bibfnamefont {C.}~\bibnamefont {Sirtori}}, \bibinfo {author} {\bibfnamefont
  {S.}~\bibnamefont {Barbieri}}, \bibinfo {author} {\bibfnamefont
  {T.}~\bibnamefont {Akalin}}, \bibinfo {author} {\bibfnamefont {J.-F.}\
  \bibnamefont {Lampin}}, \bibinfo {author} {\bibfnamefont {I.}~\bibnamefont
  {Sagnes}}, \bibinfo {author} {\bibfnamefont {H.~E.}\ \bibnamefont {Beere}}, \
  and\ \bibinfo {author} {\bibfnamefont {D.~A.}\ \bibnamefont {Ritchie}},\
  }\href {\doibase 10.1063/1.3284518} {\bibfield  {journal} {\bibinfo
  {journal} {Appl. Phys. Lett.}\ }\textbf {\bibinfo {volume} {96}},\ \bibinfo
  {pages} {021108} (\bibinfo {year} {2010})}\BibitemShut {NoStop}%
\bibitem [{\citenamefont {Gellie}\ \emph {et~al.}(2010)\citenamefont {Gellie},
  \citenamefont {Barbieri}, \citenamefont {Lampin}, \citenamefont {Filloux},
  \citenamefont {Manquest}, \citenamefont {Sirtori}, \citenamefont {Sagnes},
  \citenamefont {Khanna}, \citenamefont {Linfield}, \citenamefont {Davies},
  \citenamefont {Beere},\ and\ \citenamefont {Ritchie}}]{Gellie:10}%
  \BibitemOpen
  \bibfield  {author} {\bibinfo {author} {\bibfnamefont {P.}~\bibnamefont
  {Gellie}}, \bibinfo {author} {\bibfnamefont {S.}~\bibnamefont {Barbieri}},
  \bibinfo {author} {\bibfnamefont {J.-F.}\ \bibnamefont {Lampin}}, \bibinfo
  {author} {\bibfnamefont {P.}~\bibnamefont {Filloux}}, \bibinfo {author}
  {\bibfnamefont {C.}~\bibnamefont {Manquest}}, \bibinfo {author}
  {\bibfnamefont {C.}~\bibnamefont {Sirtori}}, \bibinfo {author} {\bibfnamefont
  {I.}~\bibnamefont {Sagnes}}, \bibinfo {author} {\bibfnamefont {S.~P.}\
  \bibnamefont {Khanna}}, \bibinfo {author} {\bibfnamefont {E.~H.}\
  \bibnamefont {Linfield}}, \bibinfo {author} {\bibfnamefont {A.~G.}\
  \bibnamefont {Davies}}, \bibinfo {author} {\bibfnamefont {H.}~\bibnamefont
  {Beere}}, \ and\ \bibinfo {author} {\bibfnamefont {D.}~\bibnamefont
  {Ritchie}},\ }\href {\doibase 10.1364/OE.18.020799} {\bibfield  {journal}
  {\bibinfo  {journal} {Opt. Express}\ }\textbf {\bibinfo {volume} {18}},\
  \bibinfo {pages} {20799} (\bibinfo {year} {2010})}\BibitemShut {NoStop}%
\bibitem [{\citenamefont {Calvar}\ \emph {et~al.}(2013)\citenamefont {Calvar},
  \citenamefont {Amanti}, \citenamefont {St-Jean}, \citenamefont {Barbieri},
  \citenamefont {Bismuto}, \citenamefont {Gini}, \citenamefont {Beck},
  \citenamefont {Faist},\ and\ \citenamefont {Sirtori}}]{Calvar2013}%
  \BibitemOpen
  \bibfield  {author} {\bibinfo {author} {\bibfnamefont {A.}~\bibnamefont
  {Calvar}}, \bibinfo {author} {\bibfnamefont {M.~I.}\ \bibnamefont {Amanti}},
  \bibinfo {author} {\bibfnamefont {M.~R.}\ \bibnamefont {St-Jean}}, \bibinfo
  {author} {\bibfnamefont {S.}~\bibnamefont {Barbieri}}, \bibinfo {author}
  {\bibfnamefont {A.}~\bibnamefont {Bismuto}}, \bibinfo {author} {\bibfnamefont
  {E.}~\bibnamefont {Gini}}, \bibinfo {author} {\bibfnamefont {M.}~\bibnamefont
  {Beck}}, \bibinfo {author} {\bibfnamefont {J.}~\bibnamefont {Faist}}, \ and\
  \bibinfo {author} {\bibfnamefont {C.}~\bibnamefont {Sirtori}},\ }\href
  {\doibase 10.1063/1.4804370} {\bibfield  {journal} {\bibinfo  {journal}
  {Appl. Phys. Lett.}\ }\textbf {\bibinfo {volume} {102}},\ \bibinfo {pages}
  {181114} (\bibinfo {year} {2013})}\BibitemShut {NoStop}%
\bibitem [{\citenamefont {Hinkov}\ \emph {et~al.}(2016)\citenamefont {Hinkov},
  \citenamefont {Hugi}, \citenamefont {Beck},\ and\ \citenamefont
  {Faist}}]{Hinkov:16}%
  \BibitemOpen
  \bibfield  {author} {\bibinfo {author} {\bibfnamefont {B.}~\bibnamefont
  {Hinkov}}, \bibinfo {author} {\bibfnamefont {A.}~\bibnamefont {Hugi}},
  \bibinfo {author} {\bibfnamefont {M.}~\bibnamefont {Beck}}, \ and\ \bibinfo
  {author} {\bibfnamefont {J.}~\bibnamefont {Faist}},\ }\href {\doibase
  10.1364/OE.24.003294} {\bibfield  {journal} {\bibinfo  {journal} {Opt.
  Express}\ }\textbf {\bibinfo {volume} {24}},\ \bibinfo {pages} {3294}
  (\bibinfo {year} {2016})}\BibitemShut {NoStop}%
\bibitem [{\citenamefont {Hangauer}\ \emph {et~al.}(2014)\citenamefont
  {Hangauer}, \citenamefont {Spinner}, \citenamefont {Nikodem},\ and\
  \citenamefont {Wysocki}}]{Hangauer:14}%
  \BibitemOpen
  \bibfield  {author} {\bibinfo {author} {\bibfnamefont {A.}~\bibnamefont
  {Hangauer}}, \bibinfo {author} {\bibfnamefont {G.}~\bibnamefont {Spinner}},
  \bibinfo {author} {\bibfnamefont {M.}~\bibnamefont {Nikodem}}, \ and\
  \bibinfo {author} {\bibfnamefont {G.}~\bibnamefont {Wysocki}},\ }\href
  {\doibase 10.1364/OE.22.023439} {\bibfield  {journal} {\bibinfo  {journal}
  {Opt. Express}\ }\textbf {\bibinfo {volume} {22}},\ \bibinfo {pages} {23439}
  (\bibinfo {year} {2014})}\BibitemShut {NoStop}%
\bibitem [{\citenamefont {Siegman}(1986)}]{Siegman1986}%
  \BibitemOpen
  \bibfield  {author} {\bibinfo {author} {\bibfnamefont {A.~E.}\ \bibnamefont
  {Siegman}},\ }\href@noop {} {\emph {\bibinfo {title} {Lasers}}}\ (\bibinfo
  {publisher} {University Science Books},\ \bibinfo {year} {1986})\BibitemShut
  {NoStop}%
\bibitem [{\citenamefont {Adler}(1973)}]{AdlerOct.}%
  \BibitemOpen
  \bibfield  {author} {\bibinfo {author} {\bibfnamefont {R.}~\bibnamefont
  {Adler}},\ }\href@noop {} {\bibfield  {journal} {\bibinfo  {journal}
  {Proceedings of the IEEE}\ }\textbf {\bibinfo {volume} {61}},\ \bibinfo
  {pages} {1380} (\bibinfo {year} {1973})}\BibitemShut {NoStop}%
\bibitem [{\citenamefont {Piccardo}\ \emph
  {et~al.}(2018{\natexlab{b}})\citenamefont {Piccardo}, \citenamefont
  {Chevalier}, \citenamefont {Anand}, \citenamefont {Wang}, \citenamefont
  {Kazakov}, \citenamefont {Mejia}, \citenamefont {Xie}, \citenamefont
  {Lascola}, \citenamefont {Belyanin},\ and\ \citenamefont
  {Capasso}}]{Piccardo2018tunable}%
  \BibitemOpen
  \bibfield  {author} {\bibinfo {author} {\bibfnamefont {M.}~\bibnamefont
  {Piccardo}}, \bibinfo {author} {\bibfnamefont {P.}~\bibnamefont {Chevalier}},
  \bibinfo {author} {\bibfnamefont {S.}~\bibnamefont {Anand}}, \bibinfo
  {author} {\bibfnamefont {Y.}~\bibnamefont {Wang}}, \bibinfo {author}
  {\bibfnamefont {D.}~\bibnamefont {Kazakov}}, \bibinfo {author} {\bibfnamefont
  {E.~A.}\ \bibnamefont {Mejia}}, \bibinfo {author} {\bibfnamefont
  {F.}~\bibnamefont {Xie}}, \bibinfo {author} {\bibfnamefont {K.}~\bibnamefont
  {Lascola}}, \bibinfo {author} {\bibfnamefont {A.}~\bibnamefont {Belyanin}}, \
  and\ \bibinfo {author} {\bibfnamefont {F.}~\bibnamefont {Capasso}},\ }\href
  {\doibase 10.1063/1.5039611} {\bibfield  {journal} {\bibinfo  {journal}
  {Appl. Phys. Lett.}\ }\textbf {\bibinfo {volume} {113}},\ \bibinfo {pages}
  {031104} (\bibinfo {year} {2018}{\natexlab{b}})}\BibitemShut {NoStop}%
\bibitem [{\citenamefont {Tavallaee}\ \emph {et~al.}(2011)\citenamefont
  {Tavallaee}, \citenamefont {Williams}, \citenamefont {Hon}, \citenamefont
  {Itoh},\ and\ \citenamefont {Chen}}]{Tavallaee2011}%
  \BibitemOpen
  \bibfield  {author} {\bibinfo {author} {\bibfnamefont {A.~A.}\ \bibnamefont
  {Tavallaee}}, \bibinfo {author} {\bibfnamefont {B.~S.}\ \bibnamefont
  {Williams}}, \bibinfo {author} {\bibfnamefont {P.~W.~C.}\ \bibnamefont
  {Hon}}, \bibinfo {author} {\bibfnamefont {T.}~\bibnamefont {Itoh}}, \ and\
  \bibinfo {author} {\bibfnamefont {Q.-S.}\ \bibnamefont {Chen}},\ }\href
  {\doibase 10.1063/1.3648104} {\bibfield  {journal} {\bibinfo  {journal}
  {Appl. Phys. Lett.}\ }\textbf {\bibinfo {volume} {99}},\ \bibinfo {pages}
  {141115} (\bibinfo {year} {2011})}\BibitemShut {NoStop}%
\bibitem [{\citenamefont {Masini}\ \emph {et~al.}(2017)\citenamefont {Masini},
  \citenamefont {Pitanti}, \citenamefont {Baldacci}, \citenamefont {Vitiello},
  \citenamefont {Degl'Innocenti}, \citenamefont {Beere}, \citenamefont
  {Ritchie},\ and\ \citenamefont {Tredicucci}}]{Masini2017}%
  \BibitemOpen
  \bibfield  {author} {\bibinfo {author} {\bibfnamefont {L.}~\bibnamefont
  {Masini}}, \bibinfo {author} {\bibfnamefont {A.}~\bibnamefont {Pitanti}},
  \bibinfo {author} {\bibfnamefont {L.}~\bibnamefont {Baldacci}}, \bibinfo
  {author} {\bibfnamefont {M.~S.}\ \bibnamefont {Vitiello}}, \bibinfo {author}
  {\bibfnamefont {R.}~\bibnamefont {Degl'Innocenti}}, \bibinfo {author}
  {\bibfnamefont {H.~E.}\ \bibnamefont {Beere}}, \bibinfo {author}
  {\bibfnamefont {D.~A.}\ \bibnamefont {Ritchie}}, \ and\ \bibinfo {author}
  {\bibfnamefont {A.}~\bibnamefont {Tredicucci}},\ }\href@noop {} {\bibfield
  {journal} {\bibinfo  {journal} {Light: Science \& Applications}\ }\textbf
  {\bibinfo {volume} {6}},\ \bibinfo {pages} {e17054} (\bibinfo {year}
  {2017})}\BibitemShut {NoStop}%
\bibitem [{\citenamefont {Pflugl}\ \emph {et~al.}(2008)\citenamefont {Pflugl},
  \citenamefont {Belkin}, \citenamefont {Wang}, \citenamefont {Geiser},
  \citenamefont {Belyanin}, \citenamefont {Fischer}, \citenamefont {Wittmann},
  \citenamefont {Faist},\ and\ \citenamefont {Capasso}}]{Pflugl2008}%
  \BibitemOpen
  \bibfield  {author} {\bibinfo {author} {\bibfnamefont {C.}~\bibnamefont
  {Pflugl}}, \bibinfo {author} {\bibfnamefont {M.~A.}\ \bibnamefont {Belkin}},
  \bibinfo {author} {\bibfnamefont {Q.~J.}\ \bibnamefont {Wang}}, \bibinfo
  {author} {\bibfnamefont {M.}~\bibnamefont {Geiser}}, \bibinfo {author}
  {\bibfnamefont {A.}~\bibnamefont {Belyanin}}, \bibinfo {author}
  {\bibfnamefont {M.}~\bibnamefont {Fischer}}, \bibinfo {author} {\bibfnamefont
  {A.}~\bibnamefont {Wittmann}}, \bibinfo {author} {\bibfnamefont
  {J.}~\bibnamefont {Faist}}, \ and\ \bibinfo {author} {\bibfnamefont
  {F.}~\bibnamefont {Capasso}},\ }\href {\doibase 10.1063/1.3009198} {\bibfield
   {journal} {\bibinfo  {journal} {Appl. Phys. Lett.}\ }\textbf {\bibinfo
  {volume} {93}},\ \bibinfo {pages} {161110} (\bibinfo {year}
  {2008})}\BibitemShut {NoStop}%
\bibitem [{\citenamefont {Kim}\ \emph {et~al.}(2018)\citenamefont {Kim},
  \citenamefont {Jung}, \citenamefont {Jiang}, \citenamefont {Fujita},
  \citenamefont {Hitaka}, \citenamefont {Ito}, \citenamefont {Edamura},\ and\
  \citenamefont {Belkin}}]{Kim:18}%
  \BibitemOpen
  \bibfield  {author} {\bibinfo {author} {\bibfnamefont {J.~H.}\ \bibnamefont
  {Kim}}, \bibinfo {author} {\bibfnamefont {S.}~\bibnamefont {Jung}}, \bibinfo
  {author} {\bibfnamefont {Y.}~\bibnamefont {Jiang}}, \bibinfo {author}
  {\bibfnamefont {K.}~\bibnamefont {Fujita}}, \bibinfo {author} {\bibfnamefont
  {M.}~\bibnamefont {Hitaka}}, \bibinfo {author} {\bibfnamefont
  {A.}~\bibnamefont {Ito}}, \bibinfo {author} {\bibfnamefont {T.}~\bibnamefont
  {Edamura}}, \ and\ \bibinfo {author} {\bibfnamefont {M.~A.}\ \bibnamefont
  {Belkin}},\ }in\ \href {\doibase 10.1364/CLEO_SI.2018.SF3G.7} {\emph
  {\bibinfo {booktitle} {Conference on Lasers and Electro-Optics}}},\ \bibinfo
  {series and number} {\bibinfo {number} {SF3G.7}}\ (\bibinfo  {publisher}
  {OSA},\ \bibinfo {year} {2018})\BibitemShut {NoStop}%
\end{thebibliography}%


\begin{thebibliography}{7}%
\makeatletter
\providecommand \@ifxundefined [1]{%
 \@ifx{#1\undefined}
}%
\providecommand \@ifnum [1]{%
 \ifnum #1\expandafter \@firstoftwo
 \else \expandafter \@secondoftwo
 \fi
}%
\providecommand \@ifx [1]{%
 \ifx #1\expandafter \@firstoftwo
 \else \expandafter \@secondoftwo
 \fi
}%
\providecommand \natexlab [1]{#1}%
\providecommand \enquote  [1]{``#1''}%
\providecommand \bibnamefont  [1]{#1}%
\providecommand \bibfnamefont [1]{#1}%
\providecommand \citenamefont [1]{#1}%
\providecommand \href@noop [0]{\@secondoftwo}%
\providecommand \href [0]{\begingroup \@sanitize@url \@href}%
\providecommand \@href[1]{\@@startlink{#1}\@@href}%
\providecommand \@@href[1]{\endgroup#1\@@endlink}%
\providecommand \@sanitize@url [0]{\catcode `\\12\catcode `\$12\catcode
  `\&12\catcode `\#12\catcode `\^12\catcode `\_12\catcode `\%12\relax}%
\providecommand \@@startlink[1]{}%
\providecommand \@@endlink[0]{}%
\providecommand \url  [0]{\begingroup\@sanitize@url \@url }%
\providecommand \@url [1]{\endgroup\@href {#1}{\urlprefix }}%
\providecommand \urlprefix  [0]{URL }%
\providecommand \Eprint [0]{\href }%
\providecommand \doibase [0]{http://dx.doi.org/}%
\providecommand \selectlanguage [0]{\@gobble}%
\providecommand \bibinfo  [0]{\@secondoftwo}%
\providecommand \bibfield  [0]{\@secondoftwo}%
\providecommand \translation [1]{[#1]}%
\providecommand \BibitemOpen [0]{}%
\providecommand \bibitemStop [0]{}%
\providecommand \bibitemNoStop [0]{.\EOS\space}%
\providecommand \EOS [0]{\spacefactor3000\relax}%
\providecommand \BibitemShut  [1]{\csname bibitem#1\endcsname}%
\let\auto@bib@innerbib\@empty
\bibitem [{\citenamefont {Wang}\ \emph {et~al.}(2017)\citenamefont {Wang},
  \citenamefont {Schwarz}, \citenamefont {Siriani}, \citenamefont {Missaggia},
  \citenamefont {Connors}, \citenamefont {Mansuripur}, \citenamefont {Calawa},
  \citenamefont {McNulty}, \citenamefont {Nickerson}, \citenamefont {Donnelly},
  \citenamefont {Creedon},\ and\ \citenamefont {Capasso}}]{Wang2017}%
  \BibitemOpen
  \bibfield  {author} {\bibinfo {author} {\bibfnamefont {C.~A.}\ \bibnamefont
  {Wang}}, \bibinfo {author} {\bibfnamefont {B.}~\bibnamefont {Schwarz}},
  \bibinfo {author} {\bibfnamefont {D.~F.}\ \bibnamefont {Siriani}}, \bibinfo
  {author} {\bibfnamefont {L.~J.}\ \bibnamefont {Missaggia}}, \bibinfo {author}
  {\bibfnamefont {M.~K.}\ \bibnamefont {Connors}}, \bibinfo {author}
  {\bibfnamefont {T.~S.}\ \bibnamefont {Mansuripur}}, \bibinfo {author}
  {\bibfnamefont {D.~R.}\ \bibnamefont {Calawa}}, \bibinfo {author}
  {\bibfnamefont {D.}~\bibnamefont {McNulty}}, \bibinfo {author} {\bibfnamefont
  {M.}~\bibnamefont {Nickerson}}, \bibinfo {author} {\bibfnamefont {J.~P.}\
  \bibnamefont {Donnelly}}, \bibinfo {author} {\bibfnamefont {K.}~\bibnamefont
  {Creedon}}, \ and\ \bibinfo {author} {\bibfnamefont {F.}~\bibnamefont
  {Capasso}},\ }\href@noop {} {\bibfield  {journal} {\bibinfo  {journal} {IEEE
  J. Sel. Top. Quantum Electron.}\ }\textbf {\bibinfo {volume} {23}},\ \bibinfo
  {pages} {1} (\bibinfo {year} {2017})}\BibitemShut {NoStop}%
\bibitem [{\citenamefont {Chattopadhyay}\ \emph {et~al.}(1981)\citenamefont
  {Chattopadhyay}, \citenamefont {Sutradhar},\ and\ \citenamefont
  {Nag}}]{Chattopadhyay1981}%
  \BibitemOpen
  \bibfield  {author} {\bibinfo {author} {\bibfnamefont {D.}~\bibnamefont
  {Chattopadhyay}}, \bibinfo {author} {\bibfnamefont {S.}~\bibnamefont
  {Sutradhar}}, \ and\ \bibinfo {author} {\bibfnamefont {B.}~\bibnamefont
  {Nag}},\ }\href@noop {} {\bibfield  {journal} {\bibinfo  {journal} {J. Phys.
  C}\ }\textbf {\bibinfo {volume} {14}},\ \bibinfo {pages} {891} (\bibinfo
  {year} {1981})}\BibitemShut {NoStop}%
\bibitem [{\citenamefont {Galavanov}\ and\ \citenamefont
  {Siukaev}(1970)}]{galavanov1970}%
  \BibitemOpen
  \bibfield  {author} {\bibinfo {author} {\bibfnamefont {V.}~\bibnamefont
  {Galavanov}}\ and\ \bibinfo {author} {\bibfnamefont {N.}~\bibnamefont
  {Siukaev}},\ }\href@noop {} {\bibfield  {journal} {\bibinfo  {journal}
  {Physica Status Solidi (b)}\ }\textbf {\bibinfo {volume} {38}},\ \bibinfo
  {pages} {523} (\bibinfo {year} {1970})}\BibitemShut {NoStop}%
\bibitem [{\citenamefont {Piccardo}\ \emph {et~al.}(2018)\citenamefont
  {Piccardo}, \citenamefont {Kazakov}, \citenamefont {Rubin}, \citenamefont
  {Chevalier}, \citenamefont {Wang}, \citenamefont {Xie}, \citenamefont
  {Lascola}, \citenamefont {Belyanin},\ and\ \citenamefont
  {Capasso}}]{Piccardo:18}%
  \BibitemOpen
  \bibfield  {author} {\bibinfo {author} {\bibfnamefont {M.}~\bibnamefont
  {Piccardo}}, \bibinfo {author} {\bibfnamefont {D.}~\bibnamefont {Kazakov}},
  \bibinfo {author} {\bibfnamefont {N.~A.}\ \bibnamefont {Rubin}}, \bibinfo
  {author} {\bibfnamefont {P.}~\bibnamefont {Chevalier}}, \bibinfo {author}
  {\bibfnamefont {Y.}~\bibnamefont {Wang}}, \bibinfo {author} {\bibfnamefont
  {F.}~\bibnamefont {Xie}}, \bibinfo {author} {\bibfnamefont {K.}~\bibnamefont
  {Lascola}}, \bibinfo {author} {\bibfnamefont {A.}~\bibnamefont {Belyanin}}, \
  and\ \bibinfo {author} {\bibfnamefont {F.}~\bibnamefont {Capasso}},\ }\href
  {\doibase 10.1364/OPTICA.5.000475} {\bibfield  {journal} {\bibinfo  {journal}
  {Optica}\ }\textbf {\bibinfo {volume} {5}},\ \bibinfo {pages} {475} (\bibinfo
  {year} {2018})}\BibitemShut {NoStop}%
\bibitem [{\citenamefont {Grover}(2004)}]{Grover2004}%
  \BibitemOpen
  \bibfield  {author} {\bibinfo {author} {\bibfnamefont {F.~W.}\ \bibnamefont
  {Grover}},\ }\href@noop {} {\emph {\bibinfo {title} {Inductance calculations:
  working formulas and tables}}}\ (\bibinfo  {publisher} {Courier
  Corporation},\ \bibinfo {year} {2004})\BibitemShut {NoStop}%
\bibitem [{\citenamefont {Alyabyeva}\ \emph {et~al.}(2017)\citenamefont
  {Alyabyeva}, \citenamefont {Zhukova}, \citenamefont {Belkin},\ and\
  \citenamefont {Gorshunov}}]{Alyabyeva2017}%
  \BibitemOpen
  \bibfield  {author} {\bibinfo {author} {\bibfnamefont {L.~N.}\ \bibnamefont
  {Alyabyeva}}, \bibinfo {author} {\bibfnamefont {E.~S.}\ \bibnamefont
  {Zhukova}}, \bibinfo {author} {\bibfnamefont {M.~A.}\ \bibnamefont {Belkin}},
  \ and\ \bibinfo {author} {\bibfnamefont {B.~P.}\ \bibnamefont {Gorshunov}},\
  }\href@noop {} {\bibfield  {journal} {\bibinfo  {journal} {Sci. Rep.}\
  }\textbf {\bibinfo {volume} {7}},\ \bibinfo {pages} {7360} (\bibinfo {year}
  {2017})}\BibitemShut {NoStop}%
\bibitem [{\citenamefont {Wang}\ and\ \citenamefont {Belyanin}(2015)}]{Wang15}%
  \BibitemOpen
  \bibfield  {author} {\bibinfo {author} {\bibfnamefont {Y.}~\bibnamefont
  {Wang}}\ and\ \bibinfo {author} {\bibfnamefont {A.}~\bibnamefont
  {Belyanin}},\ }\href {\doibase 10.1364/OE.23.004173} {\bibfield  {journal}
  {\bibinfo  {journal} {Opt. Express}\ }\textbf {\bibinfo {volume} {23}},\
  \bibinfo {pages} {4173} (\bibinfo {year} {2015})}\BibitemShut {NoStop}%
\end{thebibliography}%

\end{document}


\title{Supplementary Information to: Laser radio transmitter} 

\author{Marco Piccardo}
\email[]{These authors contributed equally to this work.}
\affiliation{Harvard John A. Paulson School of Engineering and Applied Sciences, Harvard University, Cambridge, MA 02138 USA}

\author{Michele Tamagnone}
\email[]{These authors contributed equally to this work.}
\affiliation{Harvard John A. Paulson School of Engineering and Applied Sciences, Harvard University, Cambridge, MA 02138 USA}

\author{Benedikt Schwarz}
\affiliation{Harvard John A. Paulson School of Engineering and Applied Sciences, Harvard University, Cambridge, MA 02138 USA}
\affiliation{Institute of Solid State Electronics, TU Wien, 1040 Vienna, Austria}

\author{Paul Chevalier}
\affiliation{Harvard John A. Paulson School of Engineering and Applied Sciences, Harvard University, Cambridge, MA 02138 USA}

\author{Noah A. Rubin}
\affiliation{Harvard John A. Paulson School of Engineering and Applied Sciences, Harvard University, Cambridge, MA 02138 USA}

\author{Yongrui Wang}
\affiliation{Department of Physics and Astronomy, Texas A\&M University, College Station, TX 77843, USA}

\author{Christine A. Wang}
\affiliation{Lincoln Laboratory, Massachusetts Institute of Technology, Lexington, Massachusetts 02420, USA}

\author{Michael K. Connors}
\affiliation{Lincoln Laboratory, Massachusetts Institute of Technology, Lexington, Massachusetts 02420, USA}

\author{Daniel McNulty}
\affiliation{Lincoln Laboratory, Massachusetts Institute of Technology, Lexington, Massachusetts 02420, USA}

\author{Alexey Belyanin}
\affiliation{Department of Physics and Astronomy, Texas A\&M University, College Station, TX 77843, USA}

\author{Federico Capasso}
\email[]{capasso@seas.harvard.edu}
\affiliation{Harvard John A. Paulson School of Engineering and Applied Sciences, Harvard University, Cambridge, MA 02138 USA}

\date{\today}

\maketitle 

\section{Materials and Methods}
\subsection{Laser and antenna design} The quantum cascade laser has a layer structure consisting of GaInAs/AlInAs lattice matched to InP; it emits at 9.0~$\mu$m and is described in more details in Ref.~\citenum{Wang2017}. The $12~\mu$m wide QCL waveguide was fabricated by reactive ion etching, followed by SiN passivation using plasma-enhanced chemical vapor deposition, sputtered Ti/Au contact deposition using a lift-off, substrate thinning to $150~\mu$m, bottom-side Ti/Au contact deposition and cleaving to a 8~mm long device. The device was soldered epi-side up with indium on a copper plate. The half-wave dipole antenna is designed for $f_B=5.5$~GHz and consists of two gold metal stripes (each being 6.5~mm long and 2~mm wide) with a 1~mm gap, lying onto a 3D-printed polyactide substrate (3~mm thick, $\varepsilon_r=2.7$). Each arm of the antenna is connected on one side to one of the two QCL top pads using wirebonds and, on the other side, to the negative connector of the current source via an inductor to minimize RF leakage (Figure~S~\ref{figS_model}c). The QCL is operated in a fundamental (1 FSR intermodal spacing) frequency comb regime at an injected current of $\sim$1.82~A (1.26$I_{th}$) driven with a low-noise current driver (Wavelength Electronics QCL LAB 2000) and with its temperature stabilized at 16$^\circ$C using a low-thermal-drift temperature controller (Wavelength Electronics TC5). In this operating condition the differential resistance of the QCL is estimated to be 1.3~$\Omega$, and the emitted optical power per facet is 40~mW.

\subsection{Microwave far-field measurements}
Far field mapping has been performed by mounting a commercial directive horn antenna (RF Elements SH-CC 5-30) on the same optical table with the QCL transmitter, at a distance of $\sim$0.9~m (16 $\lambda_B$) from the QCL assembly. The antenna has a maximum gain of 18.5~dBi, negligible return loss at 5.5~GHz and two separate ports for vertical and horizontal polarization, and it was aimed at the QCL transmitter. The latter is mounted on a motorized rotary stage, that allows mapping the far field in the horizontal plane. The optical table and other reflective surfaces nearby are covered with microwave absorbers (SFC-4 from Cuming Microwave) with less than 30~dB of reflectivity at 5.5~GHz. The output of the antenna (50~$\Omega$) is connected to a low noise preamplifier (19~dB of gain) and then to a spectrum analyzer (Agilent E4448A). Stage and data acquisition are controlled by a computer.

\subsection{Radio transmission} The voltage signal generated by an audio player is used to modulate the laser current using the external analog modulation input of the current source (Wavelength Electronics QCL LAB 2000, analog current transfer function: 0.4~A/V). The volume of the audio player is chosen to set the maximum peak-to-peak output voltage to a value (0.2~V) such that the maximum frequency deviation of the modulated QCL beat note lies within the demodulation bandwidth of the software defined radio (RTL-SDR: R820T tuner frequency capability 25~MHz-1750~MHz; RTL2832U demodulator bandwidth 200~kHz). 
A wideband FM (WFM) demodulation scheme is used. With reference to Fig.~2a of the main text, the amplifier gain is 19~dB at 5.5~GHz and the local oscillator (Hittite HMC-T2240) power is 0~dBm. 

\begin{figure*}[t]
\centering
\includegraphics[width=1\textwidth]{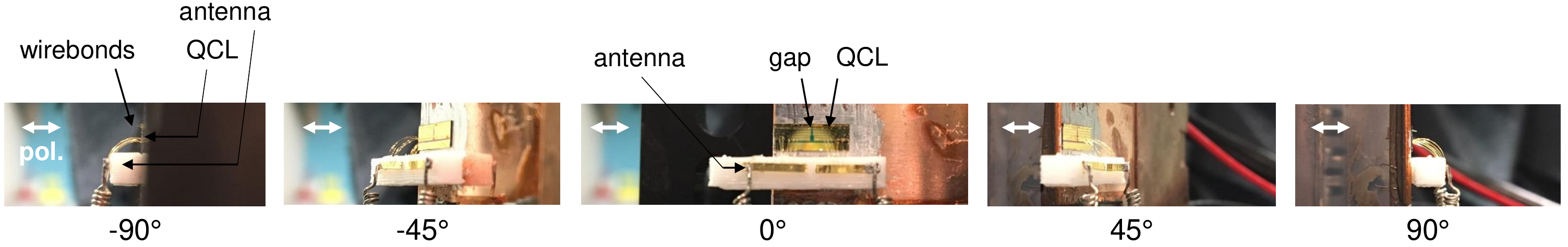}
\caption{Images of the LRT taken from the point of view of the receiving horn antenna at different rotation angles of the stage (cf. experimental set-up in Fig.2a of the main text). The direction of the polarization of the receiving horn antenna is fixed and marked as double-headed arrows in each image.}
\label{figS_rotview}
\end{figure*}

\subsection{Wireless injection locking} The master signal is generated by a local oscillator (Hittite HMC-T2240) feeding a horn antenna (RF Elements SH-CC 5-30). To monitor the changes in the QCL beat note induced by the external oscillator signal a coaxial RF probe (Quater A-20338, bandwidth DC-18 GHz, tip diameter $\sim$100~$\mu$m) is placed in the near field of the dipole antenna integrated on the QCL chip (approximately 2~mm away from the edge of one of the two arms). This arrangement is non-invasive (in the sense that the probe is not in electrical contact with the dipole antenna) and allows one to simultaneously monitor both the QCL RF beatnote and the master, radiated local oscillator signal. The signal detected by the probe is amplified (19~dB gain) and measured with a SA (Agilent E4448A).

\section{Far-field microwave radiation pattern}
The microwave radiation pattern emitted from the LRT exhibits a central peak with two side lobes (see Fig.~1d of the main text, continuous line). The central peak originates from the dipole antenna and QCL which produce a maximum close to the normal to the surface of the device (0$^\circ$), as expected for dipole-like emission patterns. The side lobes observed at angles around $-90^\circ$ and $70^\circ$ are due to emission from the wirebonds carrying RF current between the QCL and the dipole antenna. This is not obvious based on the 2D schematics shown in Fig.~1d of the main text but it becomes clear by looking at the images of the device shown in Fig.~S\ref{figS_rotview}, which are taken from the point of view of the receiving horn antenna as a function of the rotation angle of the LRT. As the dipole antenna lies on an elevated plane with respect to that of the QCL, the wirebonds carry an RF current with a significant component perpendicular to the QCL and the copper plate. The experimental configuration is such that at large rotation angles ($\pm90^\circ$, Fig.~S\ref{figS_rotview}) the wirebonds emit in the horizontal plane of study with the same polarization of the receiving antenna, thus producing the side lobes observed in Fig.~1d of the main text (continuous line). We note also that when the gap of the QCL is closed (Fig.~1d of the main text, dashed line) the RF current flowing into the wirebonds connecting the QCL with the dipole antenna is significantly reduced due to the short between the two QCL top contact pads, which almost eliminates the side lobes.

\begin{figure*}[t]
\centering
\includegraphics[width=0.6\textwidth]{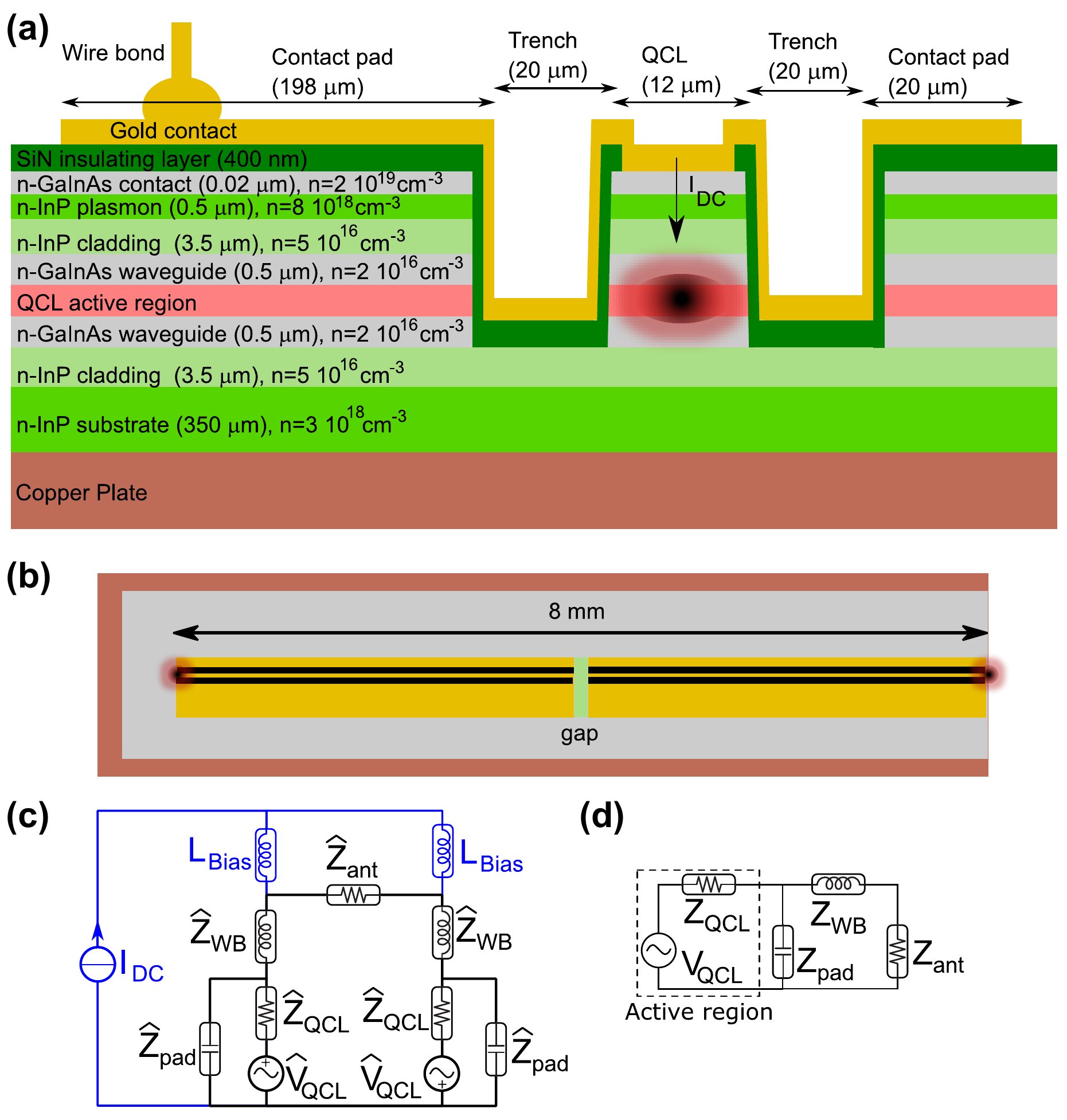}
\caption{Circuit model of the LRT. \textbf{(a)} Cross-section of the QCL with layer parameters. \textbf{(b)} Top view of the QCL with a gap in the top contact layers. \textbf{(c)} Circuit model where each half of the QCL is shown as a separate RF generator. The DC elements of the circuit are shown in blue. \textbf{(d)} Simplified RF circuit model of the LRT.}
\label{figS_model}
\end{figure*}

\section{Equivalent circuit model of the laser radio transmitter}
The cross section and top view of the laser radio transmitter (LRT) demonstrated in this work is shown in Fig.~S\ref{figS_model}a,b. Each of the sections (QCL, trenches and contact pads) can be considered as a parallel impedance that models the behavior of the layers with respect to vertical radio frequency (RF) currents, which are dominant in the system. The elements are connected in parallel by the presence of the metal contact above and the highly-doped InP substrate below. Knowing the mobility, carrier density and electrical permittivity of each layer, the admittance per unit area of each of the sections can be computed. This is done by considering the total real conductance $R_\mathrm{PUA}^{-1}$ of the layer per unit area as 

\begin{equation}
R_\mathrm{PUA}^{-1}=\frac{\sigma}{t} = \frac{n q_e \mu}{t}
\end{equation}

where $t$ is the thickness of the layer and $\sigma$ is the bulk conductivity of the material, which can be expressed as the product of the elementary charge $q_e$, the carrier density $n$ and the mobility $\mu$. For the SiN layer, $R_\mathrm{PUA}^{-1}$ is simply taken to be 0 since no charge transport occurs. Importantly, at the microwave frequency considered in this work (5.5~GHz) we neglect any effect due to the carrier inertia (plasmons) which is motivated by the fact that the frequency is well below the plasma collision frequency (an assumption that has to be revisited for terahertz frequencies).
The mobility is assumed to be 9000 cm$^2$V$^{-1}$s$^{-1}$ for GaInAs~\cite{Chattopadhyay1981} and 5000 cm$^2$V$^{-1}$s$^{-1}$ for InP~\cite{galavanov1970}.

\begin{figure*}[t]
\centering
\includegraphics[width=0.6\textwidth]{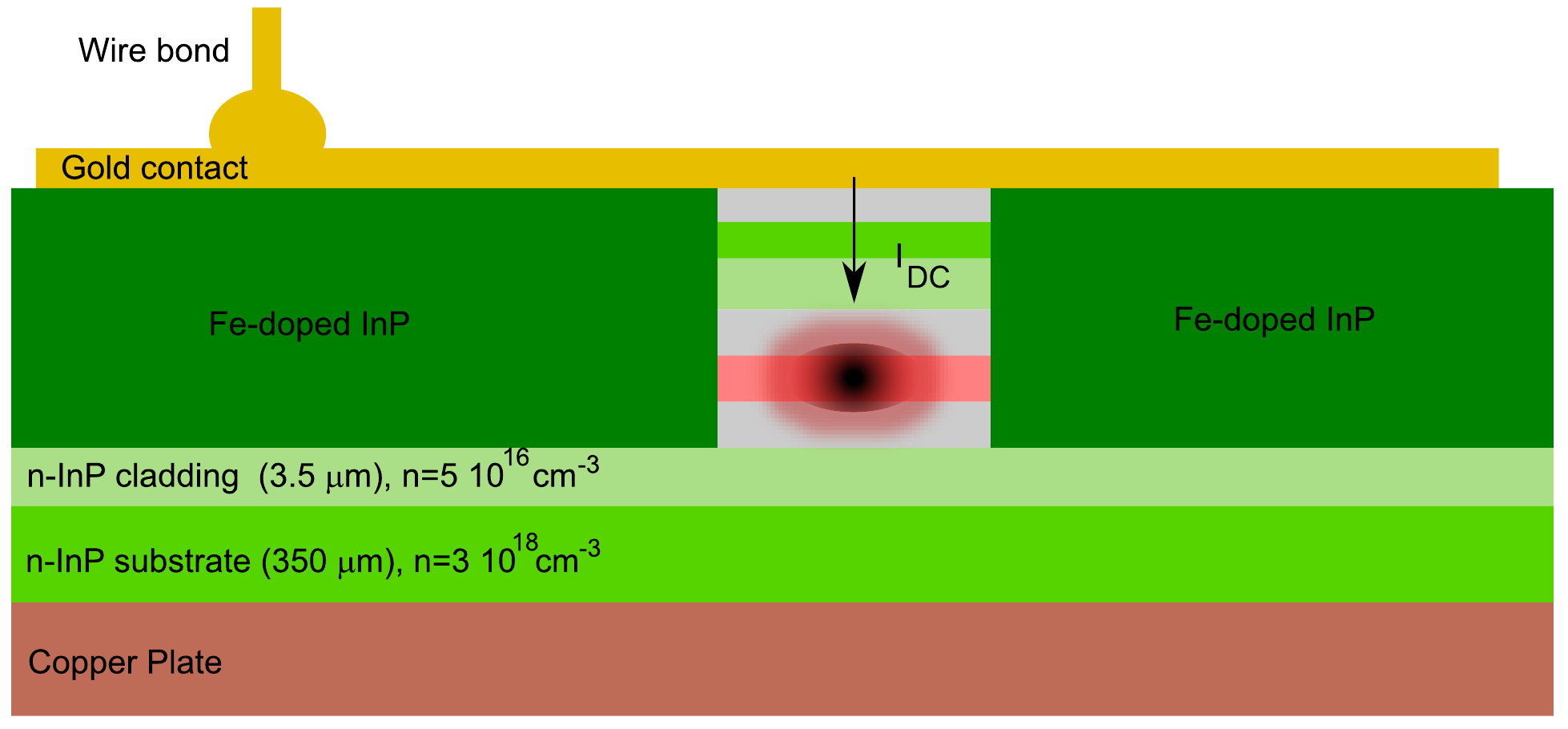}
\caption{Alternative design using Fe-doped InP to improve the impedance matching between the active region and the antenna.}
\label{figS_buried}
\end{figure*}

Secondly, we calculate the capacitance per unit area $C_\mathrm{PUA}$ of each layer as a parallel plates capacitor, namely:
\begin{equation}
C_\mathrm{PUA} = \frac{\varepsilon_0\varepsilon_r}{t}
\end{equation}

The total layer admittance is then found by taking the parallel of the real conductance and capacitance per unit area, that is:
\begin{equation}
Y_\mathrm{PUA} = R_\mathrm{PUA}^{-1} + i \omega C_\mathrm{PUA}
\end{equation}

The impedance of each layer is the reciprocal of the total admittance:
\begin{equation}
Z_\mathrm{Layer} = \frac{1}{Y_\mathrm{Layer}} = \frac{1}{A \times Y_\mathrm{PUA}} = \frac{t}{A \times (n q_e \mu + i\omega\varepsilon_0\varepsilon_r)}
\end{equation}

where $A$ is the area of the pad or trench. When several layers are stacked, the total impedance of the stack is the sum of the impedances of each layer. Finally, the parallel of the impedances of the trenches and contact pads gives, for each half of the QCL, the total pad impedance $\hat{Z}_\mathrm{pad}=0.0065 - 0.4762i~\Omega$.

\begin{figure*}[t]
\centering
\includegraphics[width=0.6\textwidth]{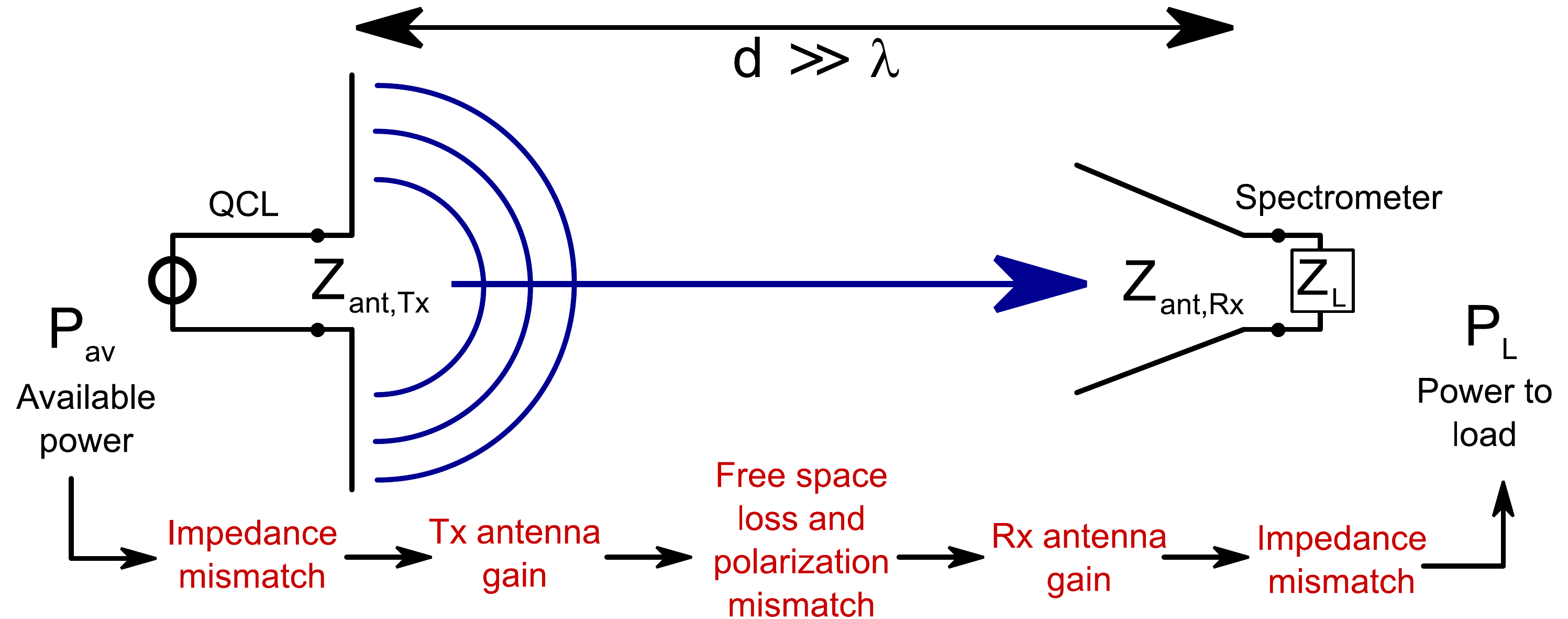}
\caption{Model of the RF wireless link between the QCL radio transmitter and the spectrum analyzer.}
\label{figS_link}
\end{figure*}

The active region of the QCL section is the source of the RF power in the system. This region is therefore represented by an RF generator including an internal impedance. In principle, both Norton equivalent circuit (based on a current generator) and Thevenin equivalent circuit (based on a voltage generator) can be used to model the source as they are interchangeable - here the Thevenin one is used. The impedance of this generator is measured directly from the derivative of the IV curve of the device. In fact, since the section with the active region is the only one that can transport DC current (due to the presence of the SiN layer in the other sections), the measured impedance at DC can only be due to the active region. The measured value is 1.3~$\Omega$ for the entire device, which is given by the parallel of the two halves of the QCL. Hence, each half of the QCL has an impedance of $\hat{Z}_\mathrm{QCL}=$2.6~$\Omega$.

The layers forming the QCL active region are present also below the contact pad, but their RF impedance is much different in the contact pad area, since no DC current is flowing through them. The impedance can be estimated using the fact that the measured low current DC impedance of the QCL active region is 493 $\Omega$. Scaling this number (that applies to the 12 $\mu$m region of the QCL) to the pad area, gives 27 $\Omega$ that is used in the model as the layer impedance $Z_\mathrm{Layer}$ for these layers.

Fig.~S\ref{figS_model}c shows the equivalent circuit model of the LRT. The two halves separated by the gap are modeled as two RF generators with opposite phases~\cite{Piccardo:18}. The pad impedance appears in parallel to the generator, and the two halves are connected to the antenna via wirebonds. The wirebonds have an inductance that can be estimated as 5.25~nH, and hence an impedance of $\hat{Z}_\mathrm{WB}=i\omega L/N=25.9i~\Omega$, where $N=7$ is the number of wire bonds on each side~\cite{Grover2004}. The antenna, instead, has an estimated impedance of $\hat{Z}_\mathrm{ant}=50~\Omega$.

When no gap in the QCL is present, the generators are shorted, meaning that very little power is radiated out of the structure (in principle none if the inductance and resistance of the top gold layer and bottom copper layer is ignored). When the gap is opened, the two parts of the circuit are now separated, and energy can be extracted using each side as an electrode of a new generator.

Because the two generators have opposite phases, from the RF point of view they are in series, and thus they can be combined in a single generator with doubled voltages and impedances (Fig.~S\ref{figS_model}d). This implies that ${Z}_\mathrm{pad}=2\hat{Z}_\mathrm{pad}=0.0130 - 0.9524i~\Omega$, ${Z}_\mathrm{QCL}=2\hat{Z}_\mathrm{QCL}=$5.2~$\Omega$ and ${Z}_\mathrm{WB}=2\hat{Z}_\mathrm{WB}=51.8i~\Omega$. Instead the impedance of the antenna is unchanged: ${Z}_\mathrm{ant}=\hat{Z}_\mathrm{ant} = 50~\Omega$. Concerning the voltage of the generator, this implies ${V}_\mathrm{QCL}=2\hat{V}_\mathrm{QCL}$, however the value used here is not important as it cancels out in the calculation, as we explain below.

The QCL has an available RF power (i.e. the maximum power that can be extracted from the generator if perfect impedance matching is realized) that can be found using the following formula:
\begin{equation}
P_\mathrm{av} = \frac{V_\mathrm{QCL}^2}{4\mathrm{Re}(Z_\mathrm{QCL})}
\label{eq_Pav}
\end{equation}

By solving the circuit and finding the actual power $P_\mathrm{ant}$ entering the antenna, it is possible to determine the ratio $P_\mathrm{ant}/P_\mathrm{av}$ independently from the voltage of the generator. The calculated ratio is in our case -22 dB and it indicates the mismatch loss between the generator and the antenna. The mismatch can be easily reduced to -9~dB if a buried QCL geometry is used (Fig.~S\ref{figS_buried}) with iron-doped InP as a passivation layer~\cite{Alyabyeva2017}. This is possible thanks to the increased thickness with respect to the SiN layer, while keeping at the same time a very good thermal conductance to dissipate the heat. The matching could be further improved by using a larger number of wirebonds and ad-hoc antennas or impedance matching networks.

Because the receiving antenna is impedance-matched with the spectrum analyzer, by knowing the received power ($P_\mathrm{Rx}=-80$ dBm, cf. Fig.~1d of the main text, continuous line) the available power at the QCL can be computed by first determining the transmitted power $P_\mathrm{Tx}=P_\mathrm{ant}$ via the Friis formula (see Fig.~S\ref{figS_link}):
\begin{equation}
P_\mathrm{Rx} = G_\mathrm{Tx}G_\mathrm{Rx}\left(\frac{\lambda}{4\pi d} \right)^2P_\mathrm{Tx}
\end{equation}

$G_\mathrm{Tx}$ and $G_\mathrm{Rx}$ are the gains of the transmitting and receiving antennas respectively, while the term in the parenthesis is the free-space loss. Knowing that $G_\mathrm{Rx} = 18.5$ dBi and estimating $G_\mathrm{Tx} = 6$ dBi we obtain that the available RF power in the QCL in the experimental operating conditions is -36 dBm. From this power value we can also back-calculate the voltage of the generator using Eq.~\ref{eq_Pav}, obtaining $\mathrm{V_{QCL}} = 2.3$~mV.

\begin{figure*}[t]
\centering
\includegraphics[width=.9\textwidth]{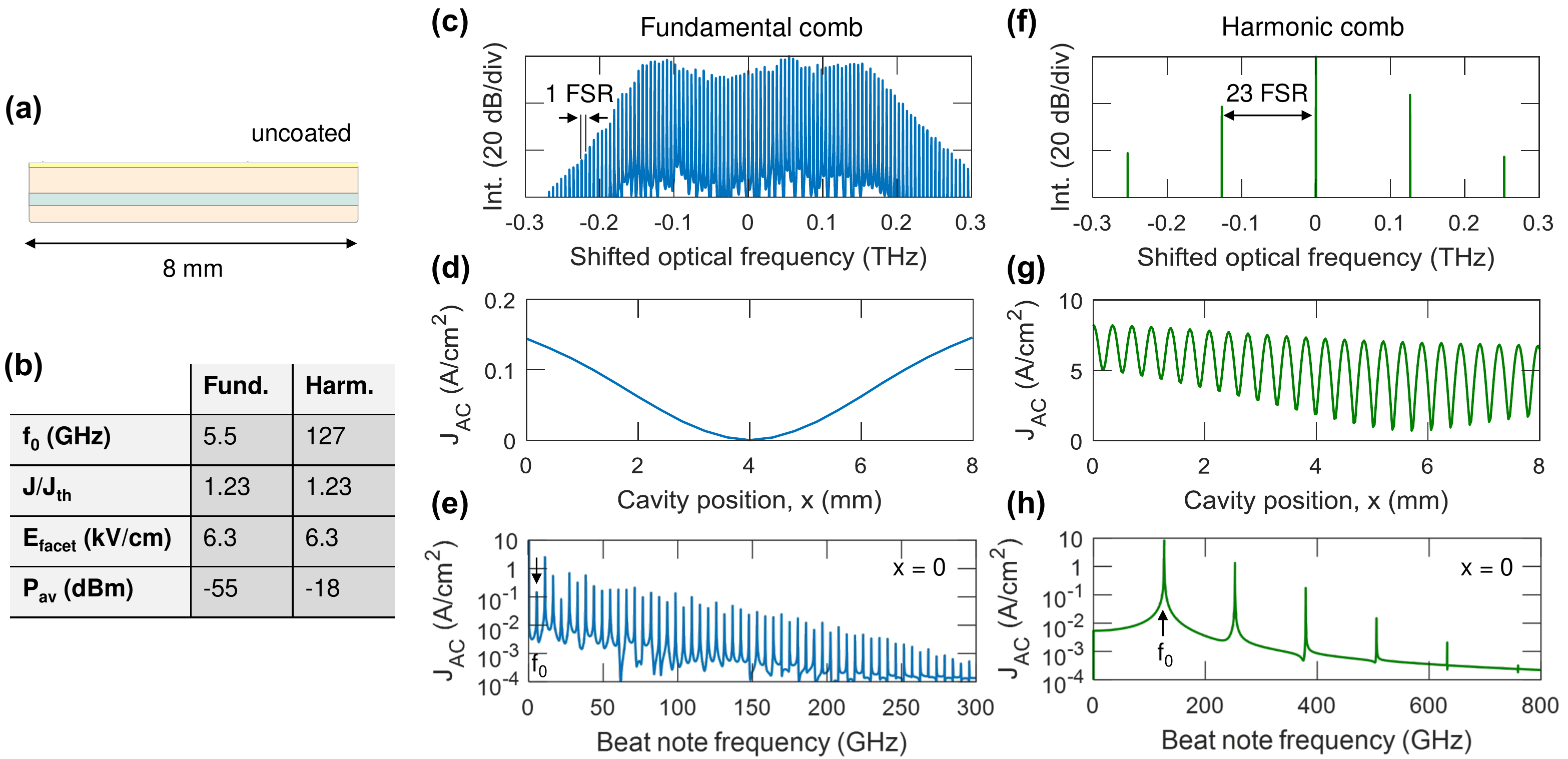}
\caption{Simulations of the QCL radio frequency generation for fundamental and harmonic comb operation. \textbf{(a)} Schematic of the uncoated Fabry-Perot cavity studied in the simulations. \textbf{(b)} Table summarizing the main simulation parameters and results. \textbf{(c)},\textbf{(f)} Optical spectra. The optical frequencies are given as relative to a central reference mode. \textbf{(d)},\textbf{(g)} Current density gratings. \textbf{(e)},\textbf{(h)} Beat note spectra calculated at the edge of the cavity ($x=0$).}
\label{figS_simulations}
\end{figure*}

\section{Simulations of the QCL radio frequency generation}
In this section we study the optical-to-radio frequency conversion of a QCL by means of space- and time-domain numerical simulations. The model we use is described in Ref.~\citenum{Wang15}. The occurrence of a dynamic population inversion gratings in numerical simulations of QCLs operating as frequency combs was alreadys shown in Ref.~\citenum{Piccardo:18}. Here we focus in particular on the difference in terms of radio frequency generation between two QCL regimes: the fundamental comb and harmonic comb.

We choose laser parameters and operating conditions representative of the mid-infrared QCL device studied experimentally in this work. Specifically, we take from bandstructure simulations of the device a gain recovery time $T_1=0.6$~ps, a dephasing time $T_2=0.12$~ps and a dipole moment $d=2.04$~nm. The simulated device has a Fabry-Perot geometry with a 8~mm long cavity, a 12~$\mu$m wide waveguide, and is operated at a current density of 1.23$\times J_{th}$ ($J_{th}$ being the current density threshold of the device). To achieve the fundamental comb regime, where the intermodal spacing is 1~free spectral range (FSR) of the laser, the electron diffusion coefficient is set to 0. On the other hand, to attain harmonic comb operation, where the intermodal spacing is a higher harmonic of the laser FSR, a weak optical seed is injected inside the cavity of the laser operating in the single mode regime. By proper tuning of the seed frequency in the vicinity of a Fabry-Perot cavity mode, a parametric process can be triggered leading to the formation of a harmonic state.

\begin{figure}[t]
\centering
\includegraphics[width=0.3\textwidth]{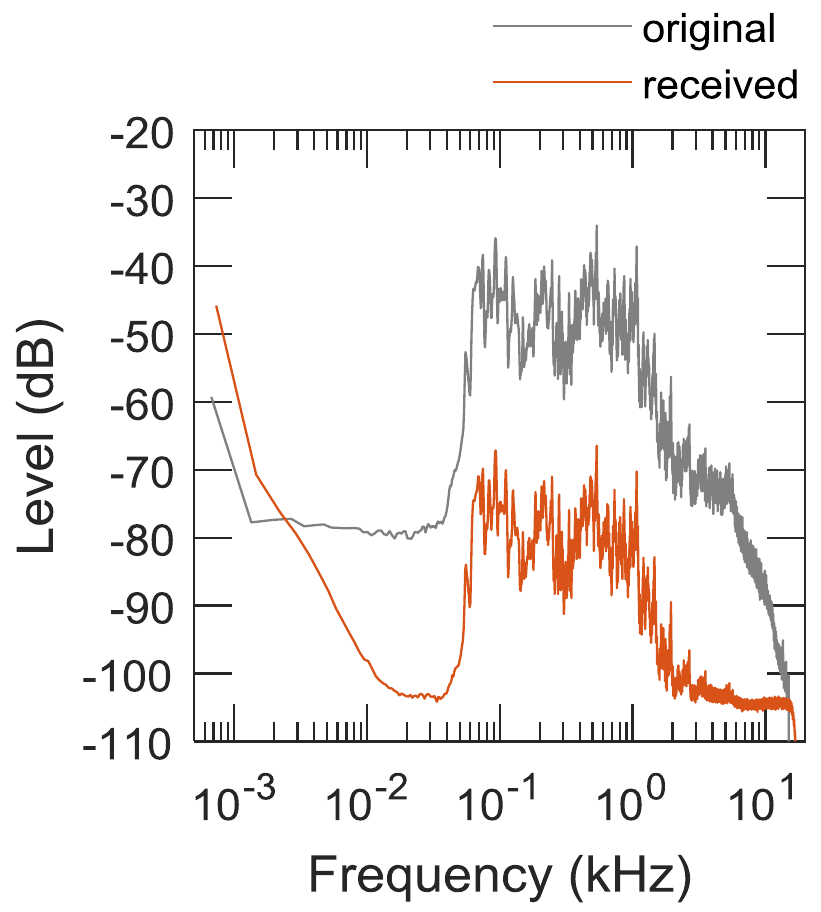}
\caption{Power spectra of the original baseband audio signal and the received one upon transmission with a quantum cascade laser radio transmitter at a carrier frequency of 5.5~GHz.}
\label{figS_audiospectra}
\end{figure}

The simulated optical spectra obtained by the laser operating in the fundamental and harmonic comb regimes are shown in Fig.~S\ref{figS_simulations}c,f, exhibiting an intermodal spacing of 5.5~GHz and 127~GHz, respectively. We note that the value of the electric field at the laser facet is very similar in the two simulations, being around 6~kV/cm. Due to the beating of the optical modes, dynamic population gratings are generated inside the cavity producing an AC component in the vertical current flowing through the laser~\cite{Piccardo:18}. The current density $J_{AC}$ gratings originating at the lowest order beat notes ($f_0$) of the laser in the fundamental and harmonic comb regimes are shown in Fig.~S\ref{figS_simulations}d,g. As expected, the number of spatial cycles scales with the number of skipped Fabry-Perot modes in the harmonic state. The striking difference between the two cases is that the peak value of $J_{AC}$ is over 60$\times$ larger in the harmonic state. This is due to spectral distribution of the optical power in the harmonic state, which exhibits fewer, more powerful optical modes, resulting in a reduced spectrum of strong beat note frequencies (cf. Fig.~S\ref{figS_simulations}e,h). In addition to this, the lowest order beat note of the fundamental comb is by itself among the weakest of the current oscillations spectrum of the laser (Fig.~S\ref{figS_simulations}e) due to the FM phase relationship among the optical modes of the QCL, which tends to suppress odd-order beat notes (see Supplementary Sec.~6D in Ref.~\citenum{Piccardo:18} for more details).

From the simulated current density gratings we can calculate the available radio frequency power at the QCL source as
\begin{equation}
P_\mathrm{av} =
\frac{1}{2} R_\mathrm{QCL} \left(w_\mathrm{wg} \int_0^L J_\mathrm{AC}(x) ~dx \right)^2
\label{eq_Pavsim}
\end{equation}
where $w_\mathrm{wg}$ and $L$ are the waveguide width and cavity length of the laser, and $R_\mathrm{QCL}=1.3~\Omega$ is the impedance of the QCL taken to be the same of the experimental device studied in this work. From Eq.~\ref{eq_Pavsim} and using the simulated current density gratings we obtain that the available power at the lowest order beat note of the harmonic state is 37~dB stronger than that of the fundamental comb. These insights show the strong potential of the harmonic comb for LRTs operating in the sub-THz range.

\section{Spectra of the transmitted and received audio tracks}
Fig.~S\ref{figS_audiospectra} shows the power spectrum of the original baseband audio signal and the one received after transmission with the QCL (cf. set-up in Fig.~2 of the main text). The signal obtained upon reception represents with fidelity the original signal except for the frequency region below 10~Hz, where additional noise is observed due to slow thermal jittering of the QCL beat note, and the frequency region around 10~kHz, due to the level of the noise floor at reception lying around -105~dB. Using an audio editing software (Audacity) the received signal is then amplified and the noise in the 10~kHz range is suppressed using a noise reduction algorithm. The SNR of the signal transmitted with the QCL could be improved by increasing the deviation bandwidth of the beat note frequency modulation (cf. Fig.~2b), namely by increasing the amplitude of the laser current modulation. In the present proof-of-concept demonstration we were limited by the demodulation bandwidth of the SDR hardware (RTL2832U demodulator  bandwidth  200  kHz).

\bibliography{wirelessbib}